\documentclass[fleqn,usenatbib]{mnras}

\usepackage[T1]{fontenc}

\DeclareRobustCommand{\VAN}[3]{#2}
\let\VANthebibliography\thebibliography
\def\thebibliography{\DeclareRobustCommand{\VAN}[3]{##3}\VANthebibliography}

\usepackage{graphicx}	
\usepackage{amsmath}	
\usepackage{amssymb}	

\usepackage{epsfig}
\usepackage{aas_macros}
\usepackage{natbib}
\usepackage{times}
\usepackage{graphicx, bm, xcolor}
\usepackage{verbatim}
\usepackage[all]{hypcap}
\usepackage{xspace}
\usepackage{multirow}
\usepackage{epstopdf}
\usepackage{pdflscape}
\usepackage{xspace}

\usepackage{newtxtext,newtxmath}

\newcommand{\msun}{\mbox{M$_\odot$}}

\newcommand{\gyr}{\mbox{${\rm Gyr}$}}
\newcommand{\pc}{\mbox{${\rm pc}$}}
\newcommand{\kpc}{\mbox{${\rm kpc}$}}
\newcommand{\mpc}{\mbox{${\rm Mpc}$}}

\newcommand{\feh}{\mbox{$[{\rm Fe}/{\rm H}]$}}
\newcommand{\be}{\begin{equation}}
\newcommand{\ee}{\end{equation}}
\newcommand{\bea}{\begin{eqnarray}}
\newcommand{\eea}{\end{eqnarray}}

\newcommand*\code[1]{{\fontfamily{lmtt}\selectfont #1}}

\newcommand{\emosaics}{{\sc E-MOSAICS}\xspace}
\newcommand{\mosaics}{{\sc MOSAICS}\xspace}
\newcommand{\eagle}{{\sc EAGLE}\xspace}

\newcommand{\subfind}{\code{SUBFIND}\xspace}
\newcommand{\fof}{\code{FoF}\xspace}

\title[Shapes of DM haloes from GCs in E-MOSAICS]{Constraining the shape of dark matter haloes with globular clusters}
\author[M. Reina-Campos et al.]{Marta~Reina-Campos$^{1,2}$\thanks{E-mail:  reinacampos@mcmaster.ca}, Sebastian~Trujillo-Gomez$^{3}$, Joel~L.~Pfeffer$^{4}$, \newauthor
Alison~Sills$^{1}$, Alis~J.~Deason$^{5,6}$, Robert~A.~Crain$^{7}$, and J.~M.~Diederik~Kruijssen$^{3}$\\
$^{1}$Department of Physics \& Astronomy, McMaster University, 1280 Main Street West, Hamilton, L8S 4M1, Canada\\
$^{2}$Canadian Institute for Theoretical Astrophysics (CITA), University of Toronto, 60 St George St, Toronto, M5S 3H8, Canada\\
$^{3}$Astronomisches Rechen-Institut, Zentrum f\"{u}r Astronomie der Universit\"{a}t Heidelberg, M\"{o}nchhofstra\ss e 12-14, 69120 Heidelberg, Germany\\
$^{4}$International Centre for Radio Astronomy Research (ICRAR), M468, University of Western Australia, 35 Stirling Hwy, Crawley, WA 6009, Australia\\
$^{5}$Institute for Computational Cosmology, Department of Physics, University of Durham, South Road, Durham DH1 3LE, UK\\
$^{6}$Centre for Extragalactic Astronomy, Department of Physics, University of Durham, South Road, Durham DH1 3LE, UK\\
$^{7}$Astrophysics Research Institute, Liverpool John Moores University, 146 Brownlow Hill, Liverpool L3 5RF, UK}

\date{Accepted XXX. Received YYY; in original form ZZZ}

\pubyear{2022}

\begin{document}
\label{firstpage}
\pagerange{\pageref{firstpage}--\pageref{lastpage}}
\maketitle

\begin{abstract}
We explore how diffuse stellar light and globular clusters (GCs) can be used to trace the matter distribution of their host halo using an observational methodology. For this, we use $117$ simulated dark matter (DM) haloes from the $(34.4~\rm cMpc)^3$ periodic volume of the \emosaics project. For each halo, we compare the stellar surface brightness and GC projected number density maps to the surface densities of DM and total mass. We find that the dominant structures identified in the stellar light and in the GCs correspond closely with those from the DM and total mass. Our method is unaffected by the presence of satellites and its precision improves with fainter GC samples. We recover tight relations between the profiles of stellar surface brightness and GC number density to those of the DM, suggesting that the profile of DM can be accurately recovered from the stars and GCs ($\sigma\leq0.5~$dex). We quantify the projected morphology of DM, stars and GCs, and find that the stars and GCs are more flattened than the DM. Additionally, the semi-major axes of the distribution of stars and GCs are typically misaligned by $\sim 10~$degrees from that of DM. We demonstrate that deep imaging of diffuse stellar light and GCs can place constraints on the shape, profile and orientation of their host halo. These results extend down to haloes with central galaxies $M_{\star}\geq10^{10}~\msun$, and the analysis will be applicable to future data from the \textit{Euclid}, \textit{Roman} and the \textit{Rubin} observatories

\end{abstract}

\begin{keywords}
galaxies: star clusters: general --- globular clusters: general --- dark matter --- galaxies: evolution --- galaxies: formation
\end{keywords}


\section{Introduction} \label{sec:intro}

One of the first probes of non-baryonic and non-luminous matter was obtained by observing the outskirts of galaxy clusters \citep{zwicky33,zwicky37}. Dark matter (DM) has since been found to make up most of the mass of the Universe \citep{planck20}, but its lack of emission of electromagnetic radiation implies that it can only be mapped using gravitational interactions. Most of the mass in DM haloes lies at large distances from the center, so probing the galactic outskirts beyond the extent of the visible galaxy is critical to trace the structure of the halo. 

In our current Lambda Cold Dark Matter ($\Lambda$CDM) paradigm, galaxies are built via a combination of in-situ star formation and hierarchical accretion of satellite galaxies. The relative role of these mechanisms changes with galaxy mass: the fraction of accreted mass in galaxies increases strongly with their mass \citep[e.g.][]{abadi06,genzel10,qu17,behroozi19}. In particular, the stellar component of galaxy clusters predominantly assembles via the accretion of satellites. An outcome of this hierarchical build-up is the presence of diffuse stellar light surrounding the central galaxy, which permeates the space between the galaxies within the cluster \citep[the `intracluster light', e.g.][]{contini21,montes22}. 

Since both the DM and the stars are collisionless, the large spatial extent covered by the intracluster light has led to it being posited as a viable tracer of the outer matter distribution of the host halo. Following this idea, \citet{montes19} demonstrate that the diffuse stellar light in six galaxy clusters follows more closely the mass distribution estimated from gravitational lensing compared to that inferred from X-ray observations. This striking result suggests that deep imaging might be sufficient to characterise the distribution of dark matter in galaxy clusters. Expanding on this work, \citet{alonsoasensio20} use the Cluster-\eagle simulations to suggest that the diffuse stellar light is an even a better tracer of the matter distribution than observations had previously suggested, and propose an indirect method to obtain the halo mass and its mass density profile from the radial profile of the intracluster mass density.

Along with the presence of diffuse stellar light, observations have also confirmed the presence of populations of intergalactic globular clusters (GCs) out to large distances in the galaxy cluster Abell 1689 \citep[e.g.][]{alamo-martinez13,alamo-martinez17}, in the Virgo cluster \citep[e.g.][]{durrell14}, in the Perseus cluster \citep[e.g.][]{harris20}, in the Fornax cluster \citep[e.g.][]{chaturvedi22}, and even in the M$81$/M$82$/NGC$3077$ system of galaxies \citep{chies-santos22}. When compared to the X-ray surface brightness distribution, the GCs in the Abell 1689 cluster show more substructure \citep{alamo-martinez13}, but those in Virgo show a similar distribution to the hot gas \citep{durrell14}. These intergalactic GCs are thus also likely an outcome of the hierarchical build-up of their current halo: these objects were stripped from their host galaxies as they were accreted, and so they now populate the outskirts of the halo. Together with the intracluster light, these GC populations are a relic of the formation and assembly history of their host cluster.

It has been established observationally that more massive haloes host a larger number of GCs \citep[e.g.][]{blakeslee97,peng08, spitler09,georgiev10,hudson14, harris15,harris17c,forbes18}. Together with the increasingly larger fraction of accreted mass in more massive haloes \citep[e.g.][]{behroozi19}, one could expect that in galactic systems in which GCs are very numerous, these objects can hold as much information about the matter distribution as the diffuse stellar light. 

Additionally, GCs are interesting objects from an observational perspective for a number of reasons. Firstly, their high luminosities \citep[i.e.~the peak being at $M_{V} \simeq -7.3$, e.g.][]{harris14} and their compact sizes \citep[$\sim3$--$4~\pc$, e.g.][]{krumholz19b} imply that their surface brightnesses are higher than that of the diffuse stellar light. Secondly, their compact nature makes them appear as bright point sources in extragalactic surveys, and with colour information they are easier to distinguish from other objects such as background galaxies \citep[e.g.][]{brito-silva21}. Lastly, their high luminosities mean that they can be observed further out from galaxy centres, i.e.~observations typically extend $\sim5$--$20$ times the stellar effective radius of the galaxy, depending on its mass \citep[e.g.][]{alabi16}. At those large radii, GCs are more likely to be tracing the outer matter distribution. Overall, GC populations can be easier and quicker to observe than the diffuse stellar light in their host haloes. For these reasons, in this work we examine both stellar and GC populations as possible tracers of the matter distribution of their host halo.

In addition to determining the shape of the DM halo, prior work has also explored correlations between the DM and baryonic morphologies, and their internal misalignments using numerical simulations of galaxy formation and evolution \citep[e.g.][]{deason11b,tenneti14,velliscig15a,velliscig15b,pillepich19,thob19,hill21}. These studies aim to provide a theoretical interpretation for upcoming measurements of the apparent alignment of galaxy shapes due to the gravitational lensing effect caused by the underlying matter distribution (e.g.~from the \textit{Euclid} and \textit{Rubin} observatories). Using the \eagle simulations, the stellar component of galaxies is found to be well aligned with the local mass distribution, but is often misaligned significantly with respect to the global halo \citep{velliscig15a, hill21}. This indicates that stars follow the local DM distribution, and that the orientation of the DM distribution changes from the inner to the outer halo. In contrast, the hot and diffuse gas in the halo (taken as a proxy for X-ray emitting gas) is found to be significantly misaligned relative to the local matter distribution. This give credence to the suggestion that the diffuse stellar light in haloes can be a good tracer of their matter distribution.

Building upon these previous studies, in this work we explore whether diffuse stellar light and GC populations trace the outer matter distribution of their host halo. We also examine if the DM surface density profile can be recovered from these observational tracers, and we extend the analysis of the projected morphologies of DM and stars to the GC populations, and their correlations. In the local Universe ($D<50~\mpc$), GC populations do not require deep observations to characterise, and might therefore provide a more efficient method to trace their host DM distribution. Upcoming deep and wide surveys with the \textit{Euclid}, \textit{Roman} and \textit{Rubin} observatories will allow us to use this method to reconstruct the host DM haloes of very large samples of galaxies using their stars and GCs. 

In this work, we use the $(34.4~\rm cMpc)^3$ periodic volume from the \emosaics project \citep{pfeffer18,kruijssen19a}. The combination of a sub-grid description for stellar cluster and formation with the \eagle galaxy formation model \citep{schaye15,crain15} allows us to study the formation and assembly of GC populations together with their host galaxies. For this, we select the $117$ DM haloes with central galaxies more massive than $M_{\star}\geq10^{10}~\msun$, which correspond to halo masses $M_{200}\gtrsim 4\times 10^{11}~\msun$. By considering lower-mass haloes that do not host galaxy clusters, we aim to investigate whether the agreement between the diffuse light and the matter distribution still holds at those masses.  

The simulation setup is described in Section~\ref{sec:emosaics}. We follow the methodology of \citet{montes19}, and explore whether the $2$D structures that can be identified in stellar surface brightness and GC number density maps correspond to the underlying structures in the DM or total mass distribution. The generation of the projected maps is discussed in Section~\ref{sec:methods}. We explore the similarity of structures identified in the DM, stellar and GC maps in Section~\ref{sec:tracing}, and examine how the DM surface density radial profile can be recovered from the stellar surface brightness and GC number density maps in Section~\ref{sec:rad-prof-dm}. We quantify the projected morphology of DM, stars and GCs in Section~\ref{sec:morphology}. Finally, a summary and a discussion about the future applicability of this analysis to observational data is presented in Section~\ref{sec:conclusions}.

\section{Simulations}\label{sec:emosaics}

\subsection{The E-MOSAICS project}\label{sub:emosaics}

The \emosaics project \citep[MOdelling Star Clusters Populations Assembly In Cosmological Simulations within \eagle project;][]{pfeffer18,kruijssen19a} combines a sub-grid description for stellar cluster formation and evolution with the \eagle galaxy formation model \citep[Evolution and Assembly of GaLaxies and their Environments;][]{schaye15,crain15}. By combining these models, these simulations allows us to study the self-consistent formation and evolution of stellar cluster populations alongside their host galaxies. We briefly summarize here the models describing the formation and evolution of sub-grid star clusters, and we refer the reader to \citet{pfeffer18} and \citet{kruijssen19a} for further details.
 
The \mosaics model adds a sub-grid stellar cluster formation and evolution model to the \eagle galaxy simulations. Every time a star particle is formed in our simulations, a sub-grid stellar cluster population can be spawned within it. The formation of the sub-grid cluster population is regulated by two physical ingredients: the cluster formation efficiency and the upper mass scale truncation of the \citet{schechter76} initial cluster mass function. We model these two ingredients with environmentally-dependent descriptions, in which natal environments with higher gas pressures lead to more mass forming in clusters and more massive objects \citep{kruijssen12d,reina-campos17}. These environmental descriptions have been found to be successful at reproducing the demographics of young massive clusters in the local Universe \citep{messa18,trujillo-gomez19,adamo20b,wainer22}.

Once the sub-grid clusters have formed, they evolve alongside their parent star particle. We model their mass evolution due to stellar evolution and dynamical processes. Mass loss due to stellar evolution is followed by the \eagle model \citep{wiersma09b}, and we model three dynamical disruption mechanisms. In our simulations, clusters lose mass due to sudden tidal shocks with their surrounding environment \citep{kruijssen11}, due to two-body interactions within the cluster, and due to their in-spiral towards the centre of galaxies caused by dynamical friction. Given the sub-grid nature of the stellar cluster populations, the latter mechanism is only applied in post-processing. To do so, we calculate the dynamical friction timescale for each cluster at every snapshot. If the timescale for in-spiral is shorter than their age, we consider that the cluster is completely disrupted.

The \emosaics project has been used to model the stellar cluster populations hosted by a volume-limited sample of $L^\star$ galaxies \citep[presented in][]{pfeffer18,kruijssen19a}. Additionally, it has also modelled a $(34.4~\rm cMpc)^3$ periodic volume, as featured in \citet{bastian20}. In this work, we study the stellar cluster populations and their host DM halos from the $(34.4~\rm cMpc)^3$ periodic volume.

The simulations reproduce many key properties of the observed galaxy population. Of particular interest for this project, the simulations reproduce the evolution of galaxy sizes across time \citep{furlong17}, the present-day luminosities and colours of galaxies \citep{trayford15} and the evolution of the galaxy stellar masses and their star formation rates \citep{furlong15}. Additionally, the simulations have been extensively used to characterize the distributions of DM, hot gas and stars, and their relationship to one another \citep[e.g.][]{velliscig15a,velliscig15b,thob19,hill21,hill22}.

The \emosaics simulations have been successful at reproducing the properties of old and young star clusters in the local Universe \citep{kruijssen19a,pfeffer19b,hughes20}, and in particular their radial profiles \citep{reina-campos21}. They have also allowed predictions for the high-redshift conditions for cluster formation \citep{pfeffer19a,reina-campos19,keller20b}, and they have showcased the potential of using GCs to trace the assembly history of their host galaxy \citep{kruijssen19b,hughes19,kruijssen20,pfeffer20,trujillo-gomez21}. Thus, these simulations provide an excellent framework to explore whether diffuse stellar and GC populations trace the overall matter distribution of their host halo. 

\subsection{Sample selection}\label{sub:sample}

We identify DM haloes in the periodic volume using the Friends-of-Friends algorithm \citep[\fof;][]{davis85} using a linking length of $0.2$ times the mean particle separation. Next, we associate gas and stellar particles to the nearest DM particle, and we identify gravitationally-bound susbtructures within each halo with the \subfind algorithm \citep{springel01,dolag09}. For this work, we select the DM haloes that contain central galaxies more massive than $M_{\star}\geq10^{10}~\msun$, which correspond to halo masses $M_{200}\gtrsim 4\times10^{11}~\msun$. This yields to a sample of $117$ haloes, which we show in the following figures using the stellar mass of the central galaxy in the DM halo. 

In our fiducial selection of particles, we include all the particles that are bound and unbound to the main halo (i.e. FOF group), as determined by the \fof algorithm. This selection includes all the particles in the central galaxy and in the diffuse unbound component, as well as those locked in the satellite galaxies within the halo.

\section{Projected distributions of DM, stars and GCs}\label{sec:methods}

In this section we present the projected distributions of DM, total mass, stars and GCs in our simulated sample of haloes, and we explore correlations between the observational tracers (i.e.~stars and GCs) and the matter components (i.e.~total mass and DM).

To do that, we first convert the simulated physical properties of stars and clusters into observable quantities, and then we produce the projected images for each component. We focus on producing mock observations rather than using direct physical measurements in order to mimic observational biases, and to provide a method that can be used directly on observational data. We also compute the correlation coefficient among pairs of our projected images, and describe the identification of the isodensity contours in the maps.

\subsection{Converting physical into observational quantities}
In order to mimic observational selection criteria, we need to calculate the luminosity emitted by stars and stellar clusters in different filters. To do so, we assume that our stars and clusters correspond to single stellar populations and we use the stellar and cluster ages, metallicities and masses together with the Flexible Stellar Population Synthesis (FSPS) model \citep{conroy09a,conroy10a,conroy10b,conroy10c} to calculate the absolute magnitudes. We further use the MILES spectral library \citep{sanchez-blazquez06}, as well as Padova isochrones \citep{girardi00,marigo07,marigo08}, a \citet{chabrier03} initial stellar mass function, and the default FSPS parameters. 

Since stellar populations in the galactic outskirts are more prominent in redder bands, we calculate the stellar luminosities in the \textit{SDSS} $r$-band. In the case of the stellar clusters, we calculate their absolute magnitudes in the $F336W$, $F475W$ and $F814W$ bands of the \textit{Hubble} Space Telescope (\textit{HST}) ACS filter system. We use these \textit{HST} bands as they have been widely used to explore extragalactic GC populations in different galactic environments \citep[e.g.][]{alamo-martinez13,harris14, harris16, alamo-martinez17}. These \textit{HST} bands roughly correspond to the $u$, $g$ and $i+z$ filters, respectively, in the \textit{SDSS} band system. All of these magnitudes are calculated in the AB system.

\subsection{Producing the projected maps}\label{sub:maps}

The aim of this study is to evaluate whether or not diffuse stellar and GC populations trace the overall structure of the matter distribution of their host halo. We compare the various populations in projected maps of the halo. It is worth noting that these simulated projected maps do not contain foreground nor background stars or galaxies, as would be the case in real observations.

Firstly, we create the maps that represent the \emph{true} matter distribution within each halo. For this, we consider both the DM and the total mass distribution (i.e.~the sum of DM, gas, and stars). We create these two sets of maps by selecting the gas, stellar and DM particles in the host halo. We then use the \code{pynbody.plot.image}\footnote{The documentation for \code{pynbody} can be found \href{http://pynbody.github.io/pynbody/}{here}} routine to produce the projected surface densities of each of these distributions. This routine uses a $k$--dimensional tree to interpolate and smooth the projected densities. Finally, the map corresponding to the projected total mass surface density corresponds to the additions of the stellar, gas and DM surface density images.

Next, we create the maps corresponding to the \emph{observational} tracers. We start by producing the stellar surface brightness maps. We do so by converting the absolute magnitudes in the \textit{SDSS} $r$--band to luminosities using a solar absolute magnitude of $M_{\odot}^{r-{\rm band}} = 4.65$ \citep{willmer18}. We then use the \code{pynbody.plot.image} routine to produce projected luminosity maps, which we convert to surface brightness maps by accounting for the angular area of a pixel in the image. Lastly, we keep all the pixels with surface brightness brighter than $\mu_{r-{\rm band}} < 28~\rm mag/arcsec^{2}$ to mimic observational constraints. We explore the effect of the surface brightness limit on our results in Sect.~\ref{sub:stars-surfbrigthlim}.

Finally, we create the number density maps of GC populations. To facilitate direct comparisons with observations, we select the GCs using observational constraints. Firstly, we select objects more metal-rich than $\feh>-3$\footnote{We discard extremely low-metallicity GCs due to numerical reasons, as these might be hosted by particles forming in unresolved galaxies and there is no model for Pop III stars in \eagle.}. Then, we apply an absolute magnitude cut in the $F475W$ band to select objects brighter than the peak of the GC mass function ($M\sim10^5~\msun$). This is needed because \emosaics has been found to underdisrupt young, low-mass and metal-rich stellar clusters \citep[see app.~D in][and discussion below]{kruijssen19a}. We use a magnitude cut of $M_{F457W} < -6.217$, which corresponds to an $8~\gyr$ old stellar cluster of solar metallicity \citep[for a solar absolute magnitude of $M_{\odot}^{457W} = 5.09$;][]{willmer18}. We apply the cut in the magnitudes rather than mass to mimic observational selection criteria. The resulting cluster sample also includes lower mass clusters than the peak of the mass function due to the metallicity dependence of the mass-to-light ratio. We examine the effect of the absolute magnitude cut on the results in Sect.~\ref{sub:gcs-maglim}.

\begin{figure}
\centering
\includegraphics[width=\hsize]{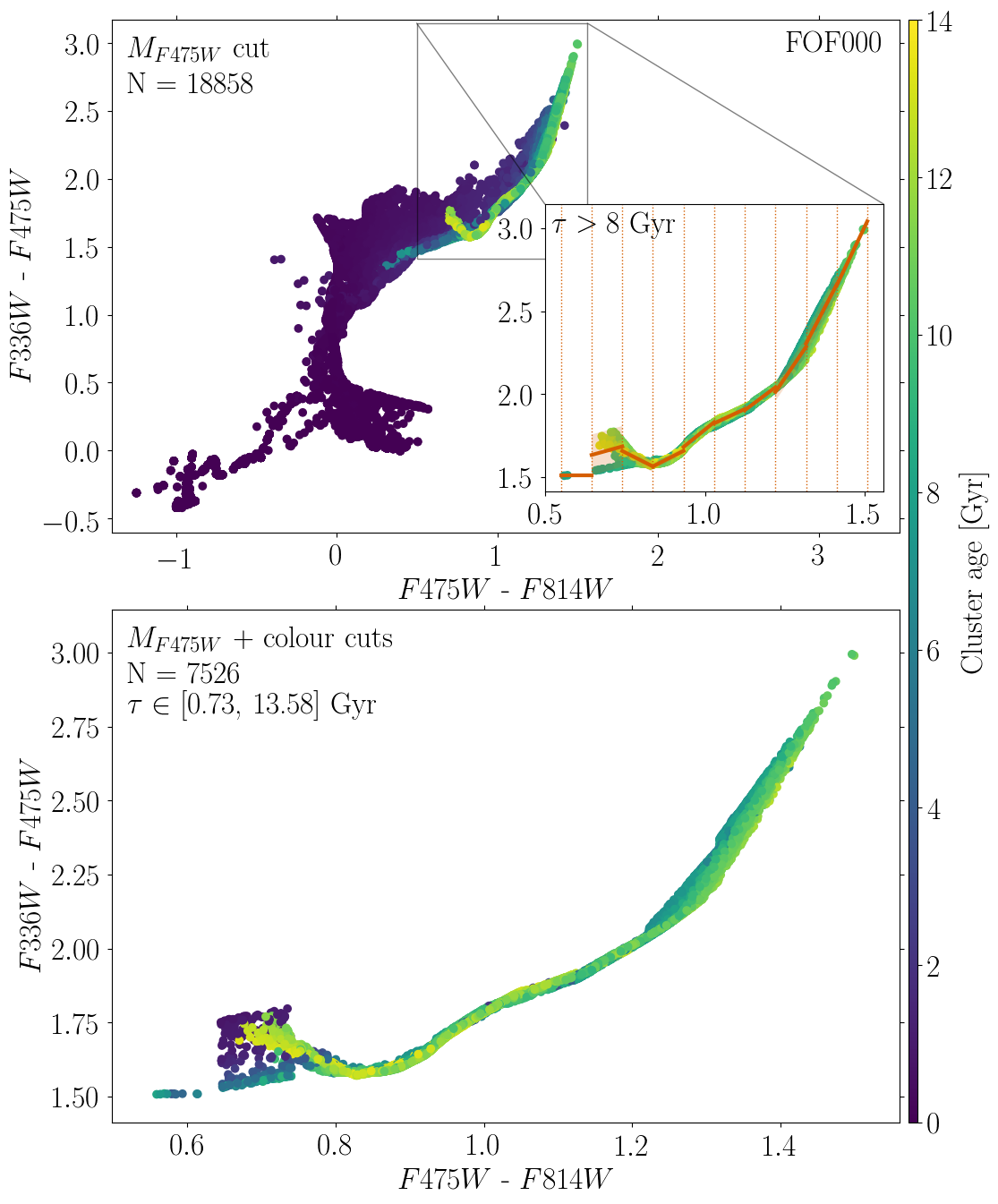}
\caption{\label{fig:gc-colour-selection} Illustration of the colour-colour selection applied to the GC populations. As an example, we show the GC population in FOF$000$. This is the most massive halo in our sample, with a mass $\log_{10}(M_{200}/\msun) = 13.7$ and a virial radius of $r_{200} = 783.8~\kpc$, and its central galaxy is a giant elliptical of stellar mass $\log_{10}(M_{\star}/\msun) = 11.3$. Panels correspond to the $F336W$--$F475W$ vs $F475W$--$F814W$ diagram of different samples of clusters. Firstly, we select stellar clusters brighter than $M_{F475W}<-6.217$ (corresponding to the peak of the GC mass function for a cluster of $8~\gyr$ and solar metallicity, \textit{top panel}). Then, we use the old clusters ($\tau>8~\gyr$) to calculate the standard deviation around linear fits performed in bins for the  $F475W$--$F814W$ colour (\textit{top panel, small inset}). Finally, we select those stellar clusters whose $F336W$--$F475W$ colour lies within $2\sigma$ of the linear fit (\textit{bottom panel}). We indicate the number of objects in each sample, and the age range of the final GC population in the upper-left corner of each panel.}
\end{figure}

\begin{figure}
\centering
\includegraphics[width=0.9\hsize]{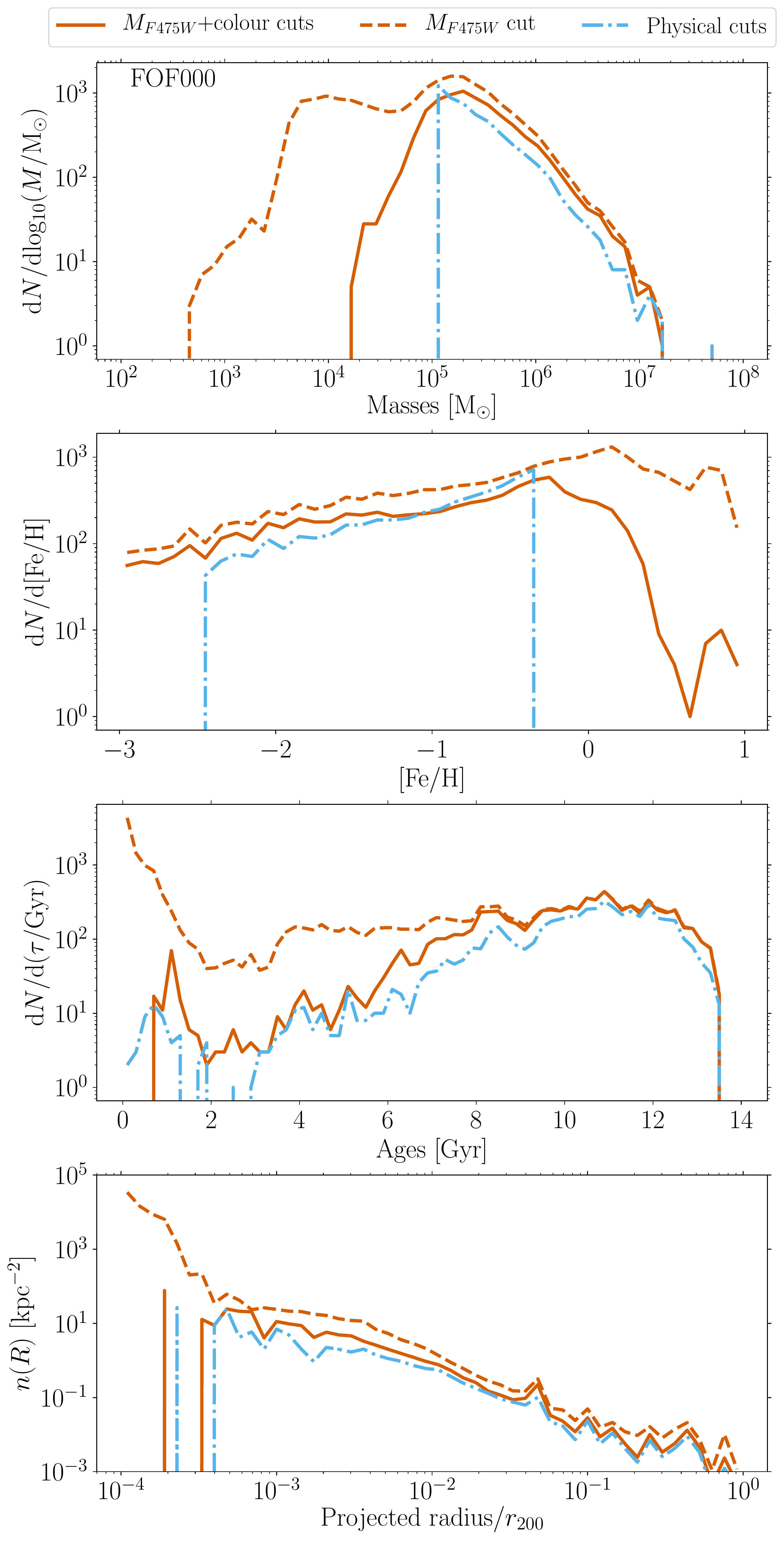}
\caption{\label{fig:hist-gc-populations} Comparison of the demographics of GC populations in FOF$000$ selected with three alternative sets of selection criteria. From top to bottom, panels correspond to the mass, metallicity, age and radial distributions, respectively. The three sets of criteria correspond to the absolute magnitude and colour-colour cut used in this work (solid red line), only an absolute magnitude cut (dashed red line) and physical cuts to select massive and metal-poor GCs (dash--dotted blue line, same selection as in \citealt{reina-campos21}). The combined use of an absolute magnitude cut in the $F475W$ band and the UV colour-colour cut is designed to be reproducible in observations while removing the low-mass, metal-rich and young stellar clusters that are artificially underdisrupted in the \emosaics simulations.}
\end{figure}

The requirement of reaching the peak of the GC luminosity function limits the distances of the DM haloes for which we will be able to apply this analysis in observational data. We can estimate these distances by considering the apparent magnitude limits expected to be reached by the \textit{Euclid} Wide Survey \citep[$m=26.2$ in the visible band;][]{scaramella21}, and the \textit{Rubin} observatory \citep[$m=27.4$ in the $g$--band after $10$ years of integration;][]{ivezic19} as these provide bracketing limits for upcoming observational facilities. For our fiducial absolute magnitude cut of $M_{F457W} < -6.217$, these observatories will be able to observe GC populations at the required completeness level in haloes up to $30.4~\mpc$ and $52.9~\mpc$ away, respectively. Given that within our simulated $(34.4~{\rm cMpc})^3$ volume there are $117$ haloes of interest, one could naively expect to find $\sim340$ and $\sim1800$ such haloes within spheres of radius $30.4~\mpc$ and $52.9~\mpc$, respectively.

As previous studies have shown, the lack of a description for the cold gas phase of the interstellar medium in the \eagle galaxy formation model implies that the low-mass, young and metal-rich stellar clusters disrupt too slowly in the \emosaics simulations \citep[see appendix D in][]{kruijssen19a}. Thus, in order to efficiently detect the contaminant objects and remove them from our samples, we use cuts in colour combinations with a UV filter ($F336W$) in addition to the cut in absolute magnitude described above. We apply the cuts in magnitude space such that a similar analysis could be carried out in observations. Identifying GCs in colour space leads to samples with more complex demographics than if simple age and metallicity cuts were applied \citep[e.g.][]{brito-silva21}. We illustrate the procedure in Fig.~\ref{fig:gc-colour-selection}. Firstly, we calculate the colours $F336W-F475W$ and $F475W-F814W$. Then, we motivate the colour cuts by examining the ranges spanned by those clusters older than $>8~\gyr$. Using $10$ bins in the $F475W-F814W$ colour, we fit a linear regression to the $F336W-F475W$ colours and we calculate the standard deviation around the fit (see small inset in the top panel of Fig.~\ref{fig:gc-colour-selection}). Lastly, we keep in our samples those clusters whose $F336W-F475W$ colour lies within $2\, \sigma$ of the linear fit. Finally, we assume that GCs are point sources and create the projected number density maps of GC populations using the \code{numpy.histogram2d} routine.

We verify that these selection criteria lead to realistic cluster populations as obtained using cuts in their physical properties in Fig.~\ref{fig:hist-gc-populations}. We look at the demographics of the selected GC populations with three alternative sets of criteria: our fiducial choice that combine an absolute magnitude in the $F475W$ band and colour-colour cuts, only an absolute magnitude cut in the $F475W$ band, and the physical cuts made to select massive and metal-poor GCs from \citet{reina-campos21}. Comparing these three GC samples in FOF$000$, we find that the combined use of an absolute magnitude cut in the $F475W$ band and the UV colour-colour cut avoids the low-mass, metal-rich and young stellar clusters that are artificially underdisrupted in the \emosaics simulations. This result applies to all the haloes in our sample.

In addition to creating the projected maps, we store the physical and observational properties of the particles used for each image. We use these data in Sect.~\ref{sec:morphology} to quantify the morphology of the projected spatial distributions of DM, stars and GCs.

\begin{figure*}
\centering
\includegraphics[width=\hsize,keepaspectratio]{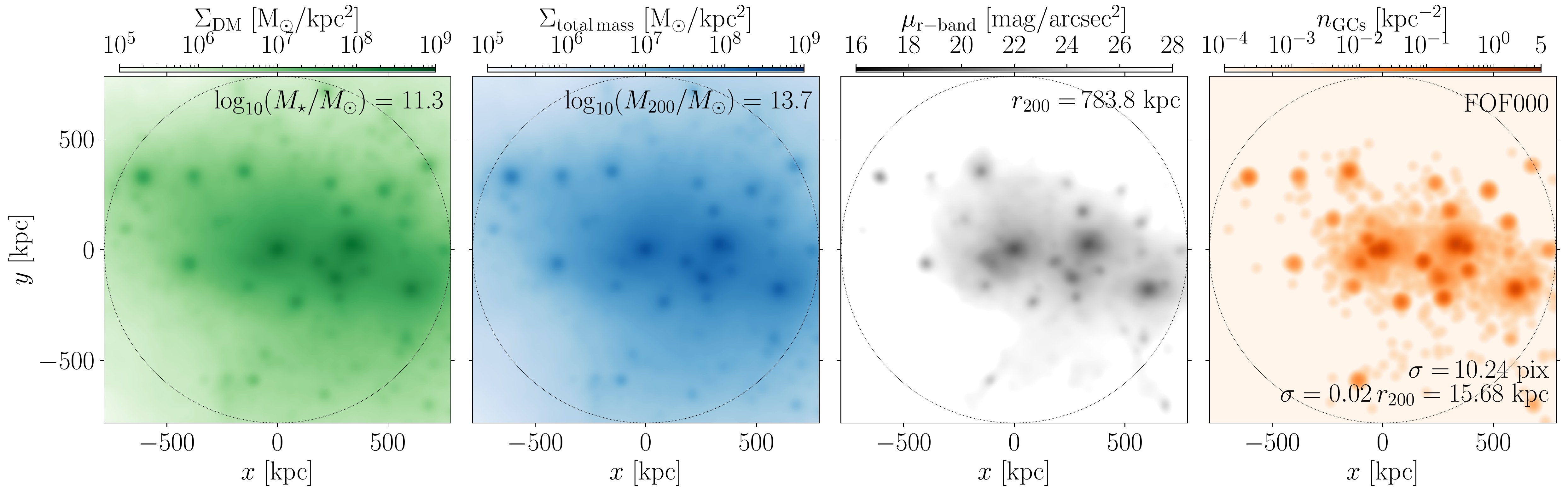}
\caption{\label{fig:proj-dm-totmass-stars-gcs-smoothed} Projected surface density maps of DM and total mass, stellar surface brightness map and number density map of GCs around FOF$000$, the most massive halo in our sample, with a mass $\log_{10}(M_{200}/\msun) = 13.7$ and a virial radius of $r_{200} = 783.8~\kpc$. Its central galaxy is a giant elliptical of stellar mass $\log_{10}(M_{\star}/\msun) = 11.3$ that seems to be undergoing a major merger at $z=0$. Stellar surface brightness maps are limited to $\mu_{r-{\rm band}} < 28~\rm mag/arcsec^{2}$ to mimic typical observational constraints. All images are smoothed by applying a Gaussian filter with a kernel size of $0.02\,r_{200}$. We indicate the physical size of the kernel in the bottom-right corner of the right-most panel. We also indicate the central galaxy stellar mass, and the virial mass and radius of the halo in the top-right corners of panels. As a reference, the thin dotted black circles mark the extent of the virial radius of the halo.}
\end{figure*}

As a last step in the production of the maps, we smooth them with a Gaussian filter of kernel size $\sigma=0.02\, r_{200}$. This scale roughly corresponds to the effective radii of the central galaxy \citep{kravtsov13}, and it removes the small scale noise from the images while preserving the large-scale structures that we aim to identify. We explore in Appendix~\ref{app:gaussian-filter} the influence of the size of the Gaussian kernel in the structures that can be identified in the images.

The resulting smoothed projected surface density maps of DM and total mass, the stellar surface brightness map, and the number density map of GCs for FOF$000$ are shown in Fig.~\ref{fig:proj-dm-totmass-stars-gcs-smoothed}. This is the most massive halo in our sample, with a mass $\log_{10}(M_{200}/\msun) = 13.7$ and a virial radius of $r_{200} = 783.8~\kpc$. Its central galaxy is a giant elliptical of stellar mass $\log_{10}(M_{\star}/\msun) = 11.3$ that seems to be undergoing a major merger at $z=0$. As expected, we find that the DM halo substructures are clearly seen in all the components. In addition, this halo also exhibits a  diffuse distribution of DM, stars and GCs around the galaxies within it. This diffuse component is a relic of the assembly history of this halo: stars and GCs are stripped during the accretion of their galaxy onto the central galaxy and left to populate the outskirts of the halo. We find similar diffuse features in all the haloes in our sample, and we aim to explore whether these diffuse stellar light and GC populations trace the matter distribution of their host halo.

\subsection{Measuring the correlation between projected maps}\label{sub:correlation}

As a first step, we determine whether the structures present in pairs of images are analogous. To do so, we measure the Pearson correlation coefficient of pixel intensities among pairs of smoothed maps\footnote{If we instead use the Spearman rank correlation coefficient, we find very similar values.}. We compute these coefficients taking as reference either the stellar surface brightness maps or the GC number density maps. We only consider those pixels in the images with significant signal, i.e.~with values above the stellar surface brightness limit or above the saturation level. Given the dynamic range of the images, we take the logarithm of the projected surface density maps of DM and of the number density maps of GCs. As a further comparison, we create fields with randomized values taken uniformly between $0$--$1$, and we compute the correlation coefficient between each of our reference maps and the random fields. 

\begin{figure}
\centering
\includegraphics[width=\hsize,keepaspectratio]{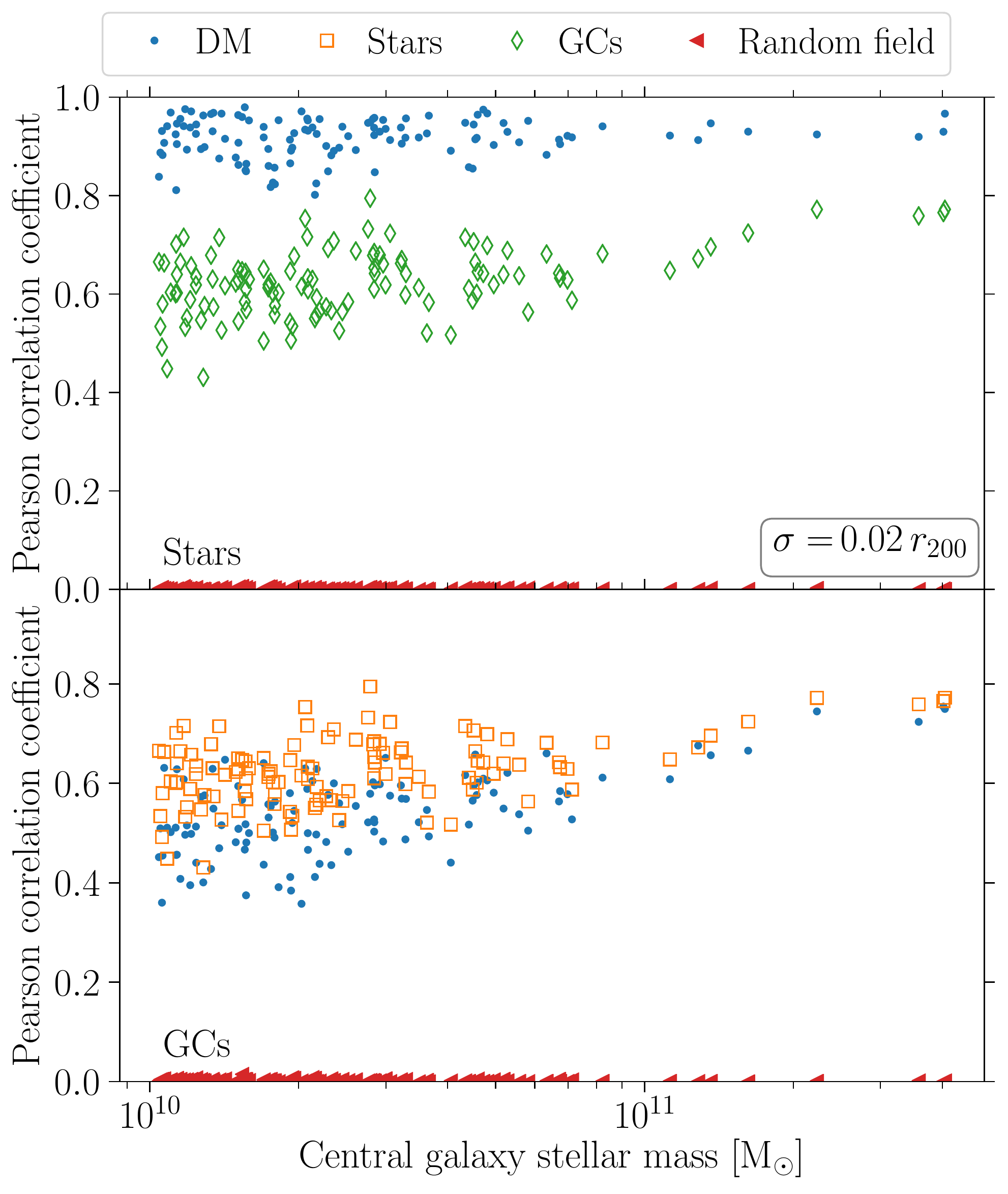}
\caption{\label{fig:pearsoncoeff-mstar} Pearson correlation coefficients between pairs of smoothed projected maps of each component as a function of the central galaxy stellar mass. We use stellar surface brightness maps as a reference in the top panel, whereas we consider projected GC number density maps in the bottom panel. We use the images smoothed with a Gaussian kernel of size $\sigma=0.02\,r_{200}$, and we only consider pixels above the background values. The random field is generated by drawing values for the pixels from a uniform distribution between $0$--$1$.}
\end{figure}

We calculate these coefficients among our sample of $117$ DM haloes, and we show the resulting values in Fig.~\ref{fig:pearsoncoeff-mstar}. Both the stellar surface brightness maps and the number density maps of GCs are uncorrelated with the randomized fields (with $p$-values of $0.4$ and $0.5$, respectively), indicating that there exists coherent structures within these images. We also find that stellar light shows the highest correlation with the underlying DM distribution, with a median coefficient of $r_{\rm P}=0.92$ across our haloes. In comparison, number density maps of GCs show a lower degree of correlation with DM (although still statistically significant), and this correlation decreases with decreasing galaxy mass. The median correlation coefficient is $r_{\rm P}=0.54$. The correlation between GCs and stars is slightly stronger, with $r_{\rm P}=0.63$. The $p$-values of all the correlations between DM, stars and GCs are consistent with being null. Thus, we find that the projected structure of the stellar and GC components shows a strong correlation with the DM.

\subsection{Identifying the isodensity contours}\label{sub:isodensity-contours}

\begin{figure*}
\centering
\includegraphics[width=\hsize,keepaspectratio]{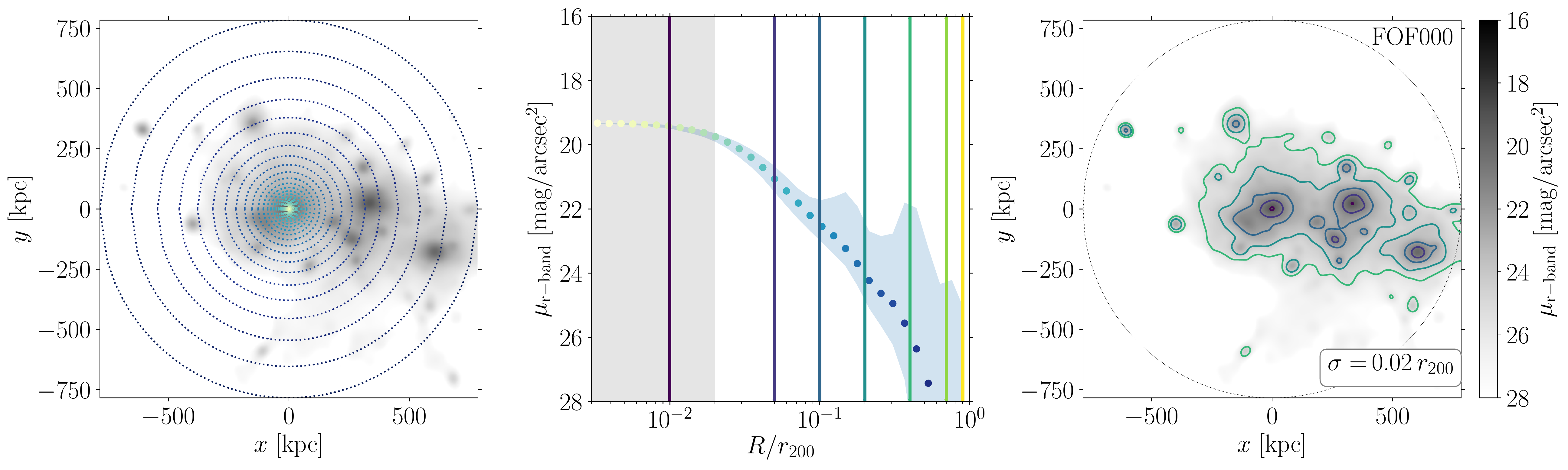}
\includegraphics[width=\hsize,keepaspectratio]{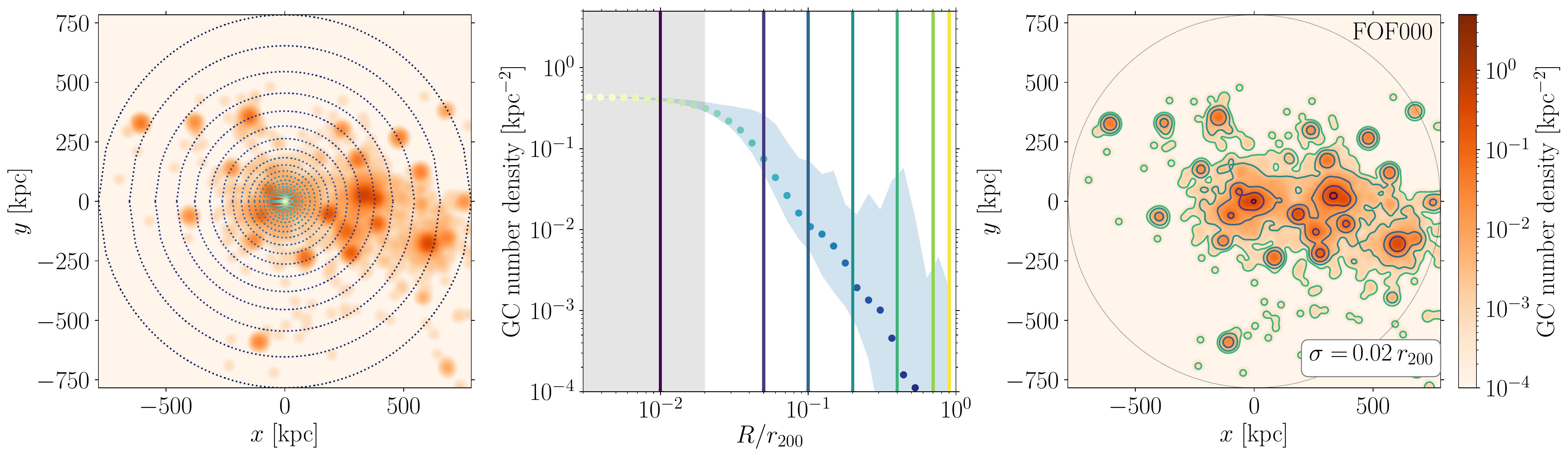}
\caption{\label{fig:id-isodensity-contours-stars-gcs} Identification of the isodensity contours in the stellar surface brightness map (\textit{top row}) and number density map of GCs (\textit{bottom row}) hosted in FOF$000$ from the \emosaics volume. The projected images are smoothed by applying a Gaussian filter with a kernel size of $2~$per cent of $r_{200}$. To identify the contours, we first calculate the median density within 32 logarithmically-spaced radial bins (\textit{left panel}), which we show as a function of the galactocentric radius (\textit{middle panel}). The blue shaded region corresponds to the $25$--$75$th percentiles, and the grey shaded region indicates the size of the Gaussian kernel. Then, we interpolate the radial profile at seven radial distances (vertical lines in the middle panel) and draw the isodensity contours corresponding to the values at those distances (\textit{right panel}). As a reference, the thin dotted black circle marks the extent of the virial radius of the halo. Structures identified with contours of the same colour correspond to the same equivalent radial bins as derived from the radial profile.}
\end{figure*}

In order to compare the structures in the projected maps of stars and GCs with those in the total mass and DM, we focus on structures defined by isodensity contours. We do this following the methodology described in \citet{montes19}. We illustrate the procedure with stellar and GC maps in Fig.~\ref{fig:id-isodensity-contours-stars-gcs}, and we provide a brief description below.

Dividing the images in 32 radial bins that are logarithmically-spaced, we calculate the median value of the density\footnote{Or surface brightness, in the case of the stellar maps.} within each bin. We interpolate the measured radial profile to obtain its value at seven radial distances, which correspond to $1,5,10,20,40,70,90~$per cent of $r_{200}$. Finally, we identify the structures in each map by joining pixels with a value equal to the interpolated value from the radial profile using the \code{matplotlib.contour} routine. At each radial distance, there can be several identified structures. Thus, we decide to use the longest contour at each level as the dominant structure. As we explore below, the presence of ongoing mergers can cause the algorithm to choose different structures in two maps for the same contour. In the next section, we explore whether the dominant isodensity contours identified in the stellar and GCs maps (i.e.~\textit{observational} distributions) correspond to the isodensity structures identified in the DM and total mass maps (i.e.~\textit{true} matter distributions).

\section{Reconstructing the shape of the DM halo}\label{sec:tracing}

In this section, we explore whether the dominant structures identified in the stellar and GC maps agree with the isodensity structures identified in the surface density maps of DM and total mass. We also explore how our results are affected by the presence of satellites in the images, the stellar surface brightness limit, and the selection criteria applied to the GC populations.

\begin{figure*}
\centering
\includegraphics[width=\hsize,keepaspectratio]{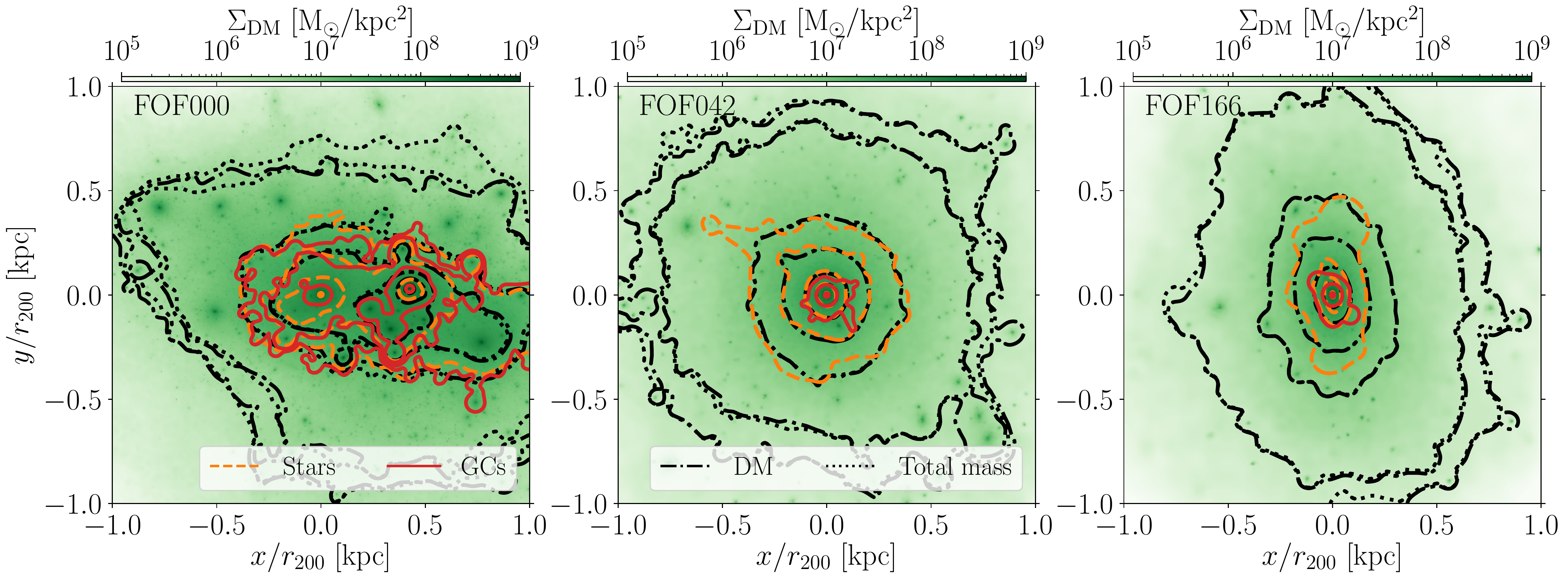}
\caption{\label{fig:overplot-dmmaps-contours} Qualitative comparison of the structures identified using isodensity contours in the surface density maps of DM and total mass (black lines), and on the stellar surface brightness maps (dashed orange lines) and the number density maps of GCs (solid red lines) for three representative haloes within the \emosaics volume. These haloes span our simulated sample with halo masses $\log_{10}(M_{200}/M_{\odot}) = 13.7, 12.2, 11.7$, respectively, and central galaxy stellar masses $\log_{10}(M_{\star}/M_{\odot}) = 11.3, 10.6, 10.0$. The isodensity contours are identified on the projected images smoothed with a Gaussian kernel size of $2~$per cent of $r_{200}$. The background images correspond to the unsmoothed surface density maps of DM in each halo. The isodensity contours can only be identified wherever there is signal in the projected image, which for lower mass haloes is generally restricted to the inner regions of the halo (e.g.~FOF$042$ and FOF$166$). }
\end{figure*}

\begin{figure*}
\centering
\includegraphics[width=\hsize,keepaspectratio]{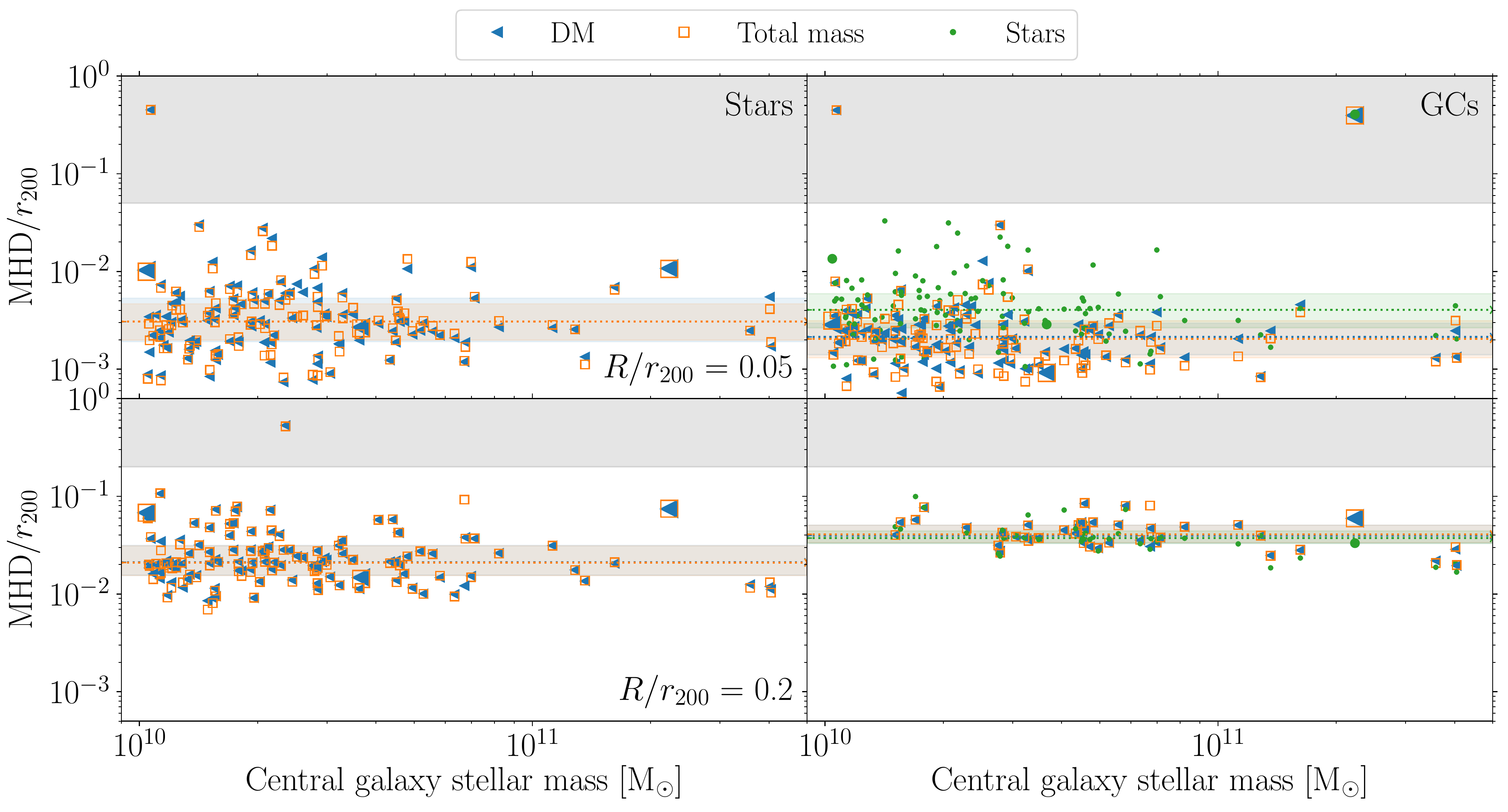}
\caption{\label{fig:mhd-r200-mstar} Modified Hausdorff Distances normalized by the size of the host halo, $r_{200}$, as a function of the central galaxy stellar mass. We calculate the MHD of isodensity contours drawn at two radial distances: $0.05\,r_{200}$ (\textit{top row}) and $0.2\,r_{200}$ (\textit{bottom row}). We consider the surface density images of DM and total mass to be the true distributions (as indicated in the legend), and we use the stellar surface brightness maps (\textit{left column}) and the number density maps of GCs (\textit{right column}) as the observational tracers. In the case of GCs, we also estimate the MHD values relative to the stellar surface brightness maps. The grey shaded regions correspond to normalized MHD larger than the radial distance at which the contour is drawn. Horizontal dotted lines with shaded regions correspond to the medians and $25$--$75$th percentiles. The three larger markers highlight the results obtained for FOF$000$, FOF$042$, and FOF$166$. GC populations are better tracers of the matter distribution within the inner halo, whereas stars are more accurate at larger radii.}
\end{figure*}

We start by doing a qualitative comparison of the isodensity contours identified in each map in three representative haloes in Fig.~\ref{fig:overplot-dmmaps-contours}. We show the isodensity structures identified in the stellar and GC maps together with those identified in the DM and total mass maps. We find that in some haloes (e.g.~FOF$000$), the presence of massive subhaloes within the halo implies that different structures are identified depending on the tracer. Based on the stellar-light contour at $0.1r_{200}$, the central galaxy in FOF$000$ dominates, but from the GC distribution we can identify a diffuse extended population associated with the prominent satellite (at $x/r_{200}, y/r_{200} \simeq 0.4, 0$). We find that ongoing major mergers can heavily distort the identification of the isodensity contours, as the incoming massive subhalo can break the assumption of smooth contours. Interestingly, these cases are very rare among our sample of simulated haloes at $z=0$. In the majority of haloes that we examine (e.g.~FOF$042$ and FOF$166$), the isodensity contours identified from the stellar and GC maps follow closely that of the DM and total mass distributions. As we examine lower mass haloes, we find that the stellar and GC distributions are restricted to the inner regions, i.e.~the isodensity contours can only be drawn for the inner $3-5$ radial distances. This is related to these haloes having undergone lower levels of satellite accretion in their assembly histories than more massive haloes, and so have formed less extended haloes.

\subsection{The Modified Hausdorff Distance}

We now perform a quantitative analysis of the similarity between the isodensity contours identified in the projected maps of each component. For that, we use the Modified Hausdorff Distance \citep[MHD;][]{huttenlocher93,dubuisson94}, which is a measure of how far apart two subsets are. This metric is a modification of the Hausdorff Distance that prevents the contamination from outliers at large distances, and it is commonly used in shape matching. For two sets of points, $X = \{x_1, x_2, ..., x_N\}$ and $Y = \{y_1, y_2, ..., y_N\}$, the MHD is defined as
\be 
{\rm MHD} (X, Y) = \max \left(d(X,Y), d(Y,X)\right),
\ee
where the distance $d(X, Y)$ is calculated as
\be
d(X, Y) = \dfrac{1}{N_{X}}\sum_{x_i \in X} \min_{y_j \in Y} || x_{i} - y_{j}||,
\ee
and it corresponds to the closest distance from a point on the contour $X$ to the contour $Y$, averaged over all points on the contour $X$.

We consider that the surface density maps of DM and total mass are the \emph{true} distributions that we aim to unveil, whereas the stellar surface brightness maps and the number density maps of GCs correspond to the \emph{observational} distributions. Hence, we calculate the MHD between pairs of isodensity contours identified in observational and true distributions among our sample of galaxies, which we show in Fig.~\ref{fig:mhd-r200-mstar}. We focus on the contours drawn at $0.05$ and $0.2\,r_{200}$ to have sufficient haloes with both stellar and GC contours to enable meaningful comparisons. We quantify the agreement between the observational and true isodensity contours with the median and $25$--$75$th percentiles, i.e.~a smaller median MHD means that the observational isodensity contour is more accurate, but a smaller interquartile range indicates that the observational tracer is more reliable overall.

\begin{figure*}
\centering
\includegraphics[width=\hsize]{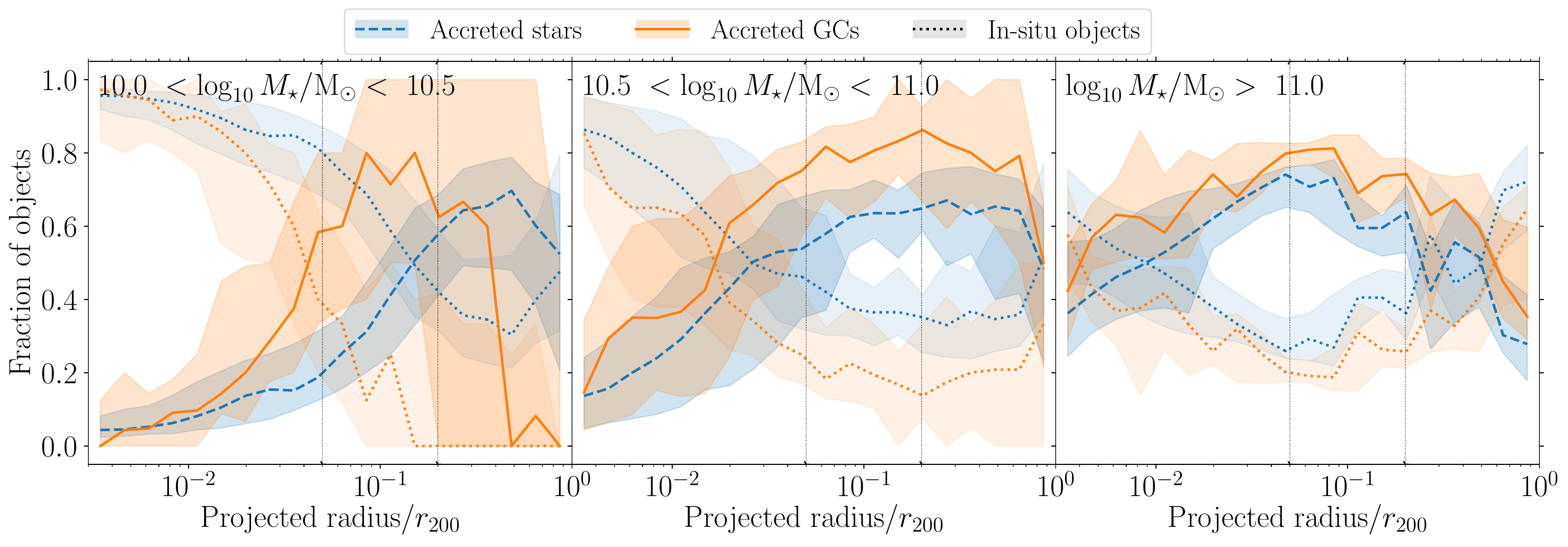}
\caption{\label{fig:accreted-fraction} Fraction of accreted stars and GCs as a function of projected radius normalized by the size of the DM halo, $r_{200}$. Each panel corresponds to a different galaxy mass bin as annotated in the top left, and the lines and shaded regions correspond to the median and $25$--$75$th percentiles among the galaxies within the bin. Dotted lines with more transparent regions indicate the fraction of in-situ stars and GCs in blue and orange, respectively. Thin black vertical dotted lines mark the radial distances at which the isodensity contours are drawn. At the radial distances considered, the fraction of accreted GCs is always larger than or equal to the fraction of accreted stars.}
\end{figure*}

The first result that we find is the similarity of the distances measured to the underlying DM and total mass distributions as expected from $\Lambda$CDM. This indicates that DM is the dominant matter component in the majority of these haloes at radii greater than $5$ per cent of $r_{200}$, and hence from here on we focus on discussing the results relative to the DM distribution. As an additional comparison, we also measure the distance between the isodensity contours of GCs and stellar light (green points in the right-hand column in Fig.~\ref{fig:mhd-r200-mstar}), and find that the median is a factor of two larger than the matter components, but the percentile range is comparable. Overall, we find that there is similar agreement between the isodensity contours of GCs and stars as with those of the matter tracers.

Comparing the isodensity contours drawn at $R/r_{200} = 0.05$, we find that the stellar light contours differ from the underlying DM contours by a median $0.003\,r_{200}$ (with percentiles $0.002$--$0.005\,r_{200}$) among our sample of haloes. In contrast, the isodensity contours from GCs differ from the DM distribution by a median $0.002\,r_{200}$ (with percentiles $0.001$--$0.003\,r_{200}$). The smaller median and scatter indicates that the projected distribution of GCs is a better tracer of the underlying DM distribution than stars in the inner region of the halo. 

Additionally, we find one system in which the measured distances between the isodensity contours of stars and DM are larger than the radius at which the contour is drawn (grey shaded region in Fig.~\ref{fig:mhd-r200-mstar}), whereas there are two outliers in the distances measured to the GCs maps. This indicates that identifying structures in the stellar surface brightness maps or in the number density maps of GCs is equally likely to be adversely affected by the presence of satellites. Overall, we find that caution is necessary when applying this method to identify isodensity structures in haloes that are undergoing major mergers.

At larger radii ($R/r_{200} = 0.2$), we find that the stellar-light contours are a median $0.02\,r_{200}$ (with percentiles $0.01$--$0.03\,r_{200}$) apart from the DM structure, whereas the isodensity contours of GCs differ by a median $0.04\,r_{200}$ (with percentiles $0.03$--$0.05\,r_{200}$). This implies that GC populations are better tracers of the matter distribution within the inner halo, whereas stars are more accurate at larger radii. 

We now examine why stars and GCs trace to the matter distribution in different regions of the halo. To do so, we explore the radial profiles of the fractions of accreted stars and GCs in Fig.~\ref{fig:accreted-fraction}. As a comparison, we also show the radial profile of the fraction of in-situ objects. We find that at $0.05\,r_{200}$, the fraction of accreted GCs is larger than that of stars across the three galaxy mass bins, whereas both fractions have comparable values at $0.2\,r_{200}$. At the radial distances probed, the accreted fractions are larger than their in-situ counterparts in galaxies more massive than $\log_{10}(M_{\star}/M_{\odot}) \geq 10.5$, i.e.~the accreted stars and GCs dominate the stellar and GC populations. Hence, the accreted components of GCs and stars are the strongest tracers of the DM halo. In the lowest galaxy mass bin, the fractions of accreted and in-situ GCs are comparable at $0.05\,r_{200}$, and similarly the fraction of accreted and in-situ stars have similar values at $0.2\,r_{200}$. Thus, the signal from the accreted component is diluted and there is more scatter in the MHDs with the DM isodensity contours. Additionally, the point source nature of GC requires smearing them to unveil the underlying continuous matter distribution, and they are more affected by the Poisson noise in the outer halo where there are few tracers. In contrast, the diffuse stellar light within the halo originated during its assembly is continuous and therefore a better tracer of the matter distribution in the outer regions.

As a proof-of-concept for this methodology, in the next subsections we test whether our results are affected by three different aspects: the presence of satellites in the images, the stellar surface brightness limit and the absolute magnitude cut applied to the GC populations.

\subsection{The effect of the presence of satellites}

\begin{figure}
\centering
\includegraphics[width=\hsize,keepaspectratio]{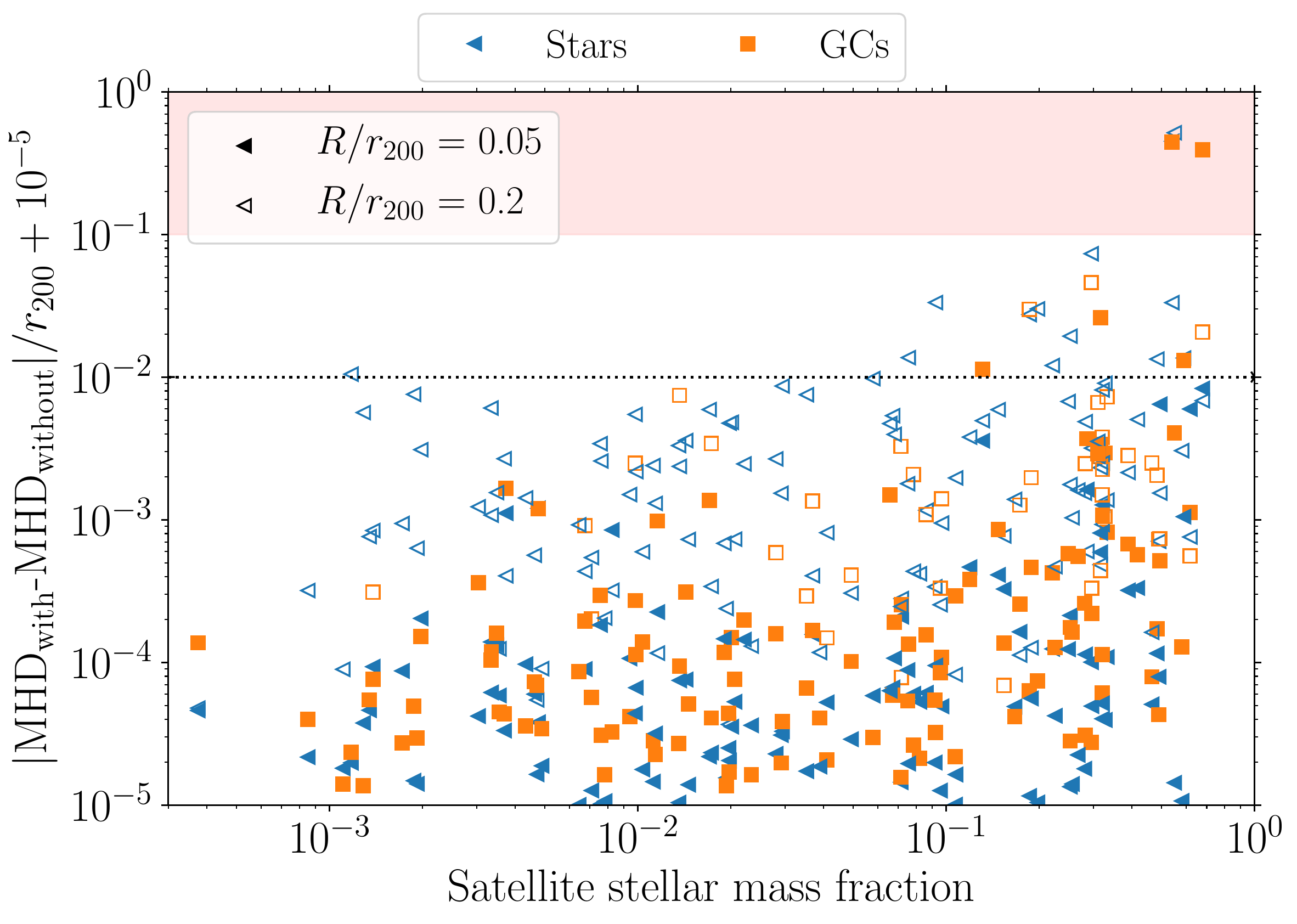}
\caption{\label{fig:compare-mhd-r200-mstar} Comparison of the Modified Hausdorff Distances to the DM contours normalized by the size of the host halo, $r_{200}$, as a function of the satellite stellar mass fraction measured in the projected images with and without substructure. We calculate the MHD of isodensity contours drawn at two radial distances: $0.05\,r_{200}$ (filled symbols) and $0.2\,r_{200}$ (empty symbols). The red shaded region indicates the parameter space in which the presence of galactic substructure affects the results by more than $10~$per cent of $r_{200}$, and the thin dotted line corresponds to $1~$per cent of $r_{200}$. The presence of satellites only affects substantially the measured MHD values for a small subset of our haloes.}
\end{figure}

In this subsection, we evaluate the effect that the presence of satellites within the haloes has on the computed MHDs. If a halo hosts a prominent satellite (e.g.~it is undergoing a major merger), the isodensity contours drawn at small radii might be more prone to picking up the satellite rather than the central galaxy in the different tracers. In that case, the longest contour used to calculate the MHD can be around two different objects (the central and the massive satellite) in the different tracers. Hence, the measured MHD would be the distance between the two galaxies, and it would be larger than expected (grey regions in Fig.~\ref{fig:mhd-r200-mstar}). We find that halo FOF$000$ is an excellent example of this issue (see Fig.~\ref{fig:overplot-dmmaps-contours}).  

In order to explore this, we consider all the particles in the main halo, and discard those that are locked in satellite galaxies. This selection is equivalent to the `diffuse mass' component from \citet{pillepich18}. This selection reflects the populations that are bound to the central galaxy, as well as the diffuse objects that lie in the main halo as a result of its accretion history. 

Following our image treatment, we smooth these maps with a Gaussian kernel of size $0.02\,r_{200}$, and identify the isodensity contours (Sect.~\ref{sub:isodensity-contours}). As in the previous section, we then compute the MHD between the isodensity contours identified in the DM and the observed maps at two radial distances. Finally, we compare the obtained MHDs to the values obtained from the images that contain substructure, which we show in Fig.~\ref{fig:compare-mhd-r200-mstar} as a function of the mass fraction in satellite galaxies within the halo. 

The distances measured from the images with and without substructure at $0.05\,r_{200}$ are overall less affected by the presence of satellites in the images than when comparing the contours drawn at $0.2\,r_{200}$. When examining haloes with increasing stellar mass locked in satellite galaxies (i.e.~with increasing satellite mass fractions), we find that there is an increasing trend: the contours drawn in the images with satellites are increasingly further apart from those without satellites in them. This trend is especially noticeable in haloes with more than $20~$per cent of their stellar mass in satellite galaxies.

There are four instances in which the distances measured from the images with and without subhaloes differ significantly (i.e.~by more than $10~$per cent of $r_{200}$). All of the four markers above $0.3$ (the two left-most markers are overlapping) correspond to the markers that lie in the grey regions of Fig.~\ref{fig:mhd-r200-mstar}. These haloes have more than $50~$per cent of their stellar mass in satellite galaxies, possibly indicating the presence of one or more massive satellites within the halo. This indicates that ongoing major mergers, or haloes with a large number of satellites, correspond to cases where the method breaks down.

In the vast majority of cases, we find that the presence of substructure in our images has an effect smaller than $0.01\,r_{200}$ for the two radial distances considered here. This implies that we can apply this methodology to observational data without the need of removing satellite galaxies from the images (unless the halo contains an ongoing major merger).

\subsection{Modifying the stellar surface brightness limit}\label{sub:stars-surfbrigthlim}

\begin{figure}
\centering
\includegraphics[width=\hsize,keepaspectratio]{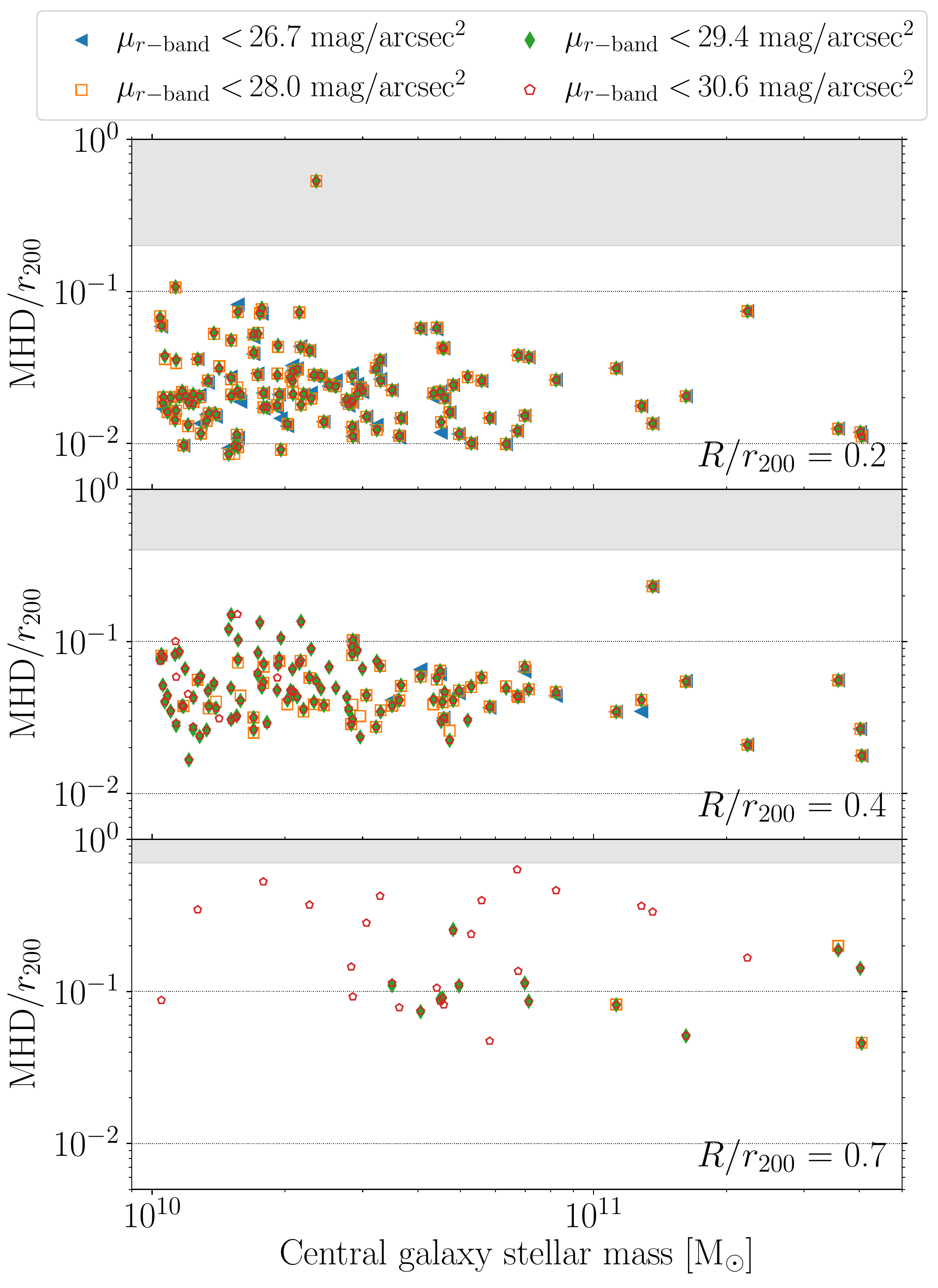}
\caption{\label{fig:compare-mhd-r200-mstar-surfbrightlim} Modified Hausdorff Distances between DM contours and stellar contours identified in maps of varying depth. The MHD is normalized by the size of the host halo, $r_{200}$, and shown as a function of the galaxy stellar mass. We calculate the MHD between DM and stellar isodensity contours drawn at three radial distances: $0.2\,r_{200}$ (\textit{top panel}), $0.4\,r_{200}$ (\textit{middle panel}) and $0.7\,r_{200}$ (\textit{bottom panel}). The grey shaded regions correspond to normalized MHD larger than the radial distance at which the contours are drawn. Deeper stellar observations allow us to apply this methodology to lower-mass haloes and to further out regions.}
\end{figure}

We examine here whether modifying the surface brightness limit of the stellar maps with substructure has an effect on the measured MHD relative to the DM isodensity contours.

To do this, we produce stellar surface brightness images at different limiting depths by masking out pixels fainter than given values. We consider the stellar surface brightness limits of \textit{SDSS}, \citep[$\mu_{r-{\rm band}} = 26.7~\rm mag/arcsec^{2}$;][]{york00}, the \textit{Euclid} Wide Survey, \citep[$\mu_{r-{\rm band}} = 29.4~\rm mag/arcsec^{2}$;][]{borlaff22} and the limit after 10 years of full survey integration from the \textit{Rubin} $r$--band, \citep[$\mu_{r-{\rm band}} = 30.6~\rm mag/arcsec^{2}$;][]{ivezic19}, in addition to our fiducial limit of $\mu_{r-{\rm band}} = 28~\rm mag/arcsec^{2}$. 

For each of these maps, we first smooth them with a Gaussian kernel of size $0.02\,r_{200}$ and identify the isobrightness contours as described in Sect.~\ref{sub:isodensity-contours}. We then compute the MHD between them and the DM isodensity contours, which we show in Fig.~\ref{fig:compare-mhd-r200-mstar-surfbrightlim}. Given that deeper observations reveal further details in the outskirts of the haloes, we calculate the MHD at three radial distances that probe the outer regions of the halo: $0.2\,r_{200}$, $0.4\,r_{200}$ and $0.7\,r_{200}$. As is expected, deepening the stellar surface brightness limit allows us to measure the stellar distribution at larger radii for more haloes. At a radius of $0.4\,r_{200}$, there are $57$ haloes in our fiducial images ($\mu_{r-{\rm band}}=28~\rm mag/arcsec^{2}$) for which we can measure the MHD, whereas we can do so for the entire sample of $117$ haloes when lowering the cut to the end-of-mission \textit{Rubin} limit. Pushing outwards to a distance of $0.7\,r_{200}$, there are $3$ and $33$ haloes for which we can measure their MHDs in our fiducial and \textit{Rubin} images, respectively. 

Interestingly, deeper observations barely increase the precision of the MHD compared to shallower images of the same halo. Hence, this implies that deeper surveys will allow us to apply this methodology to lower-mass haloes and further out in the galaxies, but they would not improve the accuracy of the reconstruction of the DM distribution using shallower data.

\subsection{Modifying the GC absolute magnitude limit}\label{sub:gcs-maglim}

\begin{figure}
\centering
\includegraphics[width=\hsize,keepaspectratio]{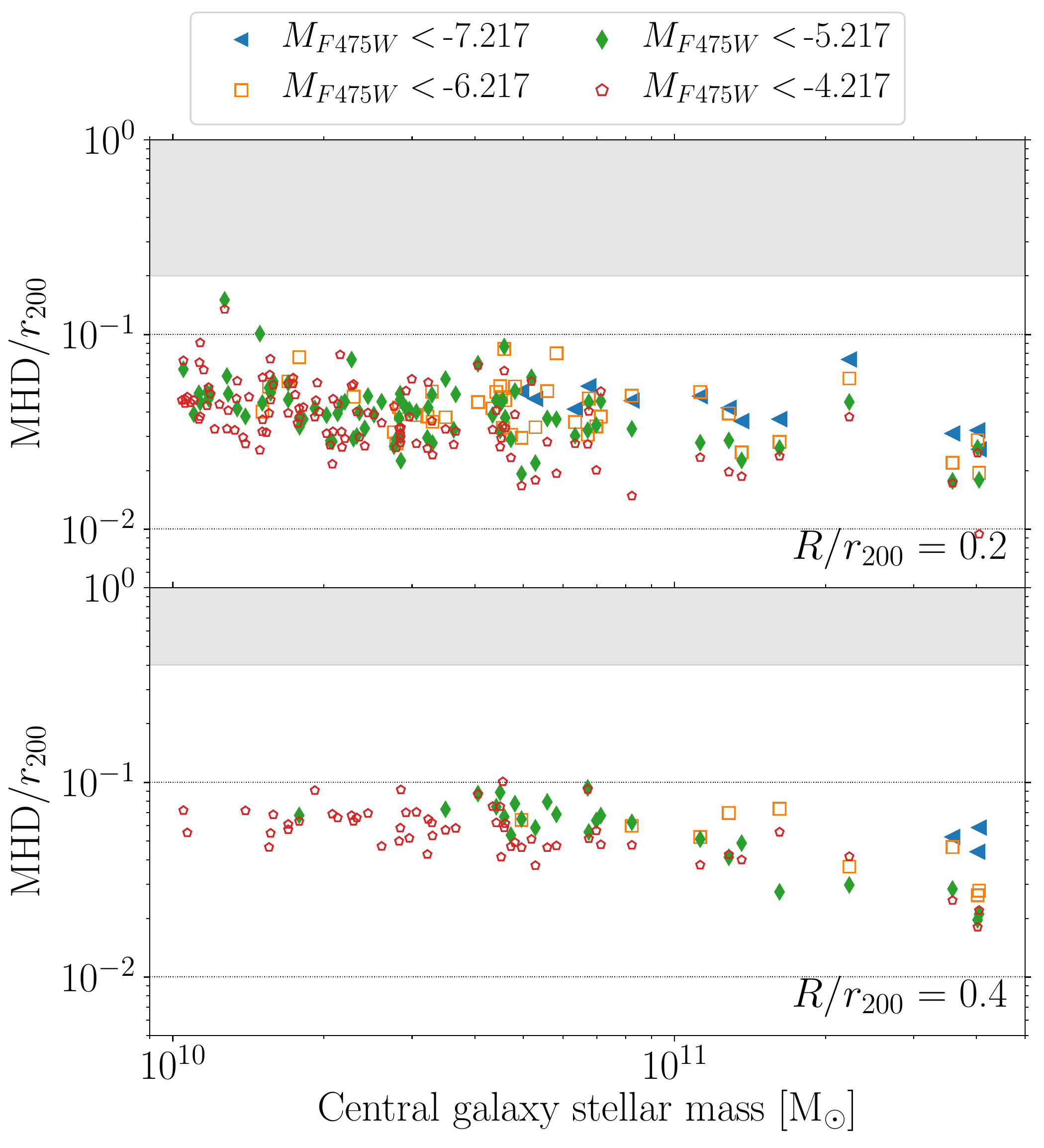}
\caption{\label{fig:compare-mhd-r200-mstar-maglim} Modified Hausdorff Distances between DM contours and contours identified over GC populations of varying depth. The MHD is normalized by the size of the host halo, $r_{200}$, and shown as a function of the central galaxy stellar mass. We calculate the MHD between DM and GC isodensity contours drawn at two radial distances: $0.2\,r_{200}$ (\textit{top panel}) and $0.4\,r_{200}$ (\textit{bottom panel}). The grey shaded regions correspond to normalized MHD larger than the radial distance at which the contours are drawn. Fainter GC populations allow us to reconstruct the DM distribution more accurately and extend the method to lower mass galaxies and larger radii.}
\end{figure}

We now examine the influence of the selection criteria of GC populations on the measured MHDs relative to the DM contours. In particular, we focus on varying the absolute magnitude limit in the $F475W$ band. The fiducial absolute magnitude limit of $M_{F475W} = -6.217$ corresponds to the peak of the GC mass function ($\sim 10^5~\msun$) for a $8~\gyr$ old cluster of solar metallicity. Thus, by increasing and decreasing this value we are selecting GC populations that contain, respectively, lower and higher-mass clusters than the peak of the mass function. This is important for predicting the accuracy of our method for reconstructing the DM distribution when applied to galaxies at different distances. Regardless of the absolute magnitude limit used, we also select GCs based on colour combinations with a UV filter to remove the contamination from underdisrupted objects (see Sect.~\ref{sub:maps}).

Using these different absolute magnitude limits, we produce the corresponding projected number density maps of the GC populations, and we smooth them with a Gaussian kernel of size $0.02\,r_{200}$. We then identify the isodensity contours (Sect.~\ref{sub:isodensity-contours}), and compute the MHD between them and the DM contours, which we show in Fig.~\ref{fig:compare-mhd-r200-mstar-maglim}. 

Observing fainter GC populations ($M_{F475W} < -4.217$) allows us to extend this analysis to further distances for more systems, and more accurately. Using the faintest GC samples, we can reconstruct the DM distribution for $114$ and $57$ haloes at $0.2\,r_{200}$ and $0.4\,r_{200}$, respectively. In comparison, there are $13$ and $3$ haloes, respectively, for which we can do the same analysis with only the brightest GC populations. Additionally, we find that observing fainter GC populations leads to a more accurate recovery of the DM distribution (i.e.~smaller MHDs). This is in contrast with the result obtained varying the stellar surface brightness limit (Sect.~\ref{sub:stars-surfbrigthlim}), in which deeper observations only added information towards the outskirts of the haloes. 

We find an increasing trend of MHDs towards lower-mass galaxies when comparing the contours drawn at $0.4\,r_{200}$ in the faintest GC populations. This trend can be related to the mass range of the satellite galaxies accreted by these haloes. Lower-mass satellites host fewer GCs \citep[e.g.][]{harris17c} and therefore contribute fewer accreted GCs that can be left to populate the outskirts during their host galaxy accretion. Lastly, we find that this analysis can only be applied at a distance of $0.7\,r_{200}$ for the faintest GC populations in the six most massive haloes. For those haloes, the measured MHDs from the GC contours are $\sim0.1\,r_{200}$ away from the DM distribution. Hence, these results suggest that GC populations can provide good estimates of the shape of the matter distribution out to $0.4\,r_{200}$ of their host DM halo for haloes with central galaxy stellar masses of $M_{\star}\gtrsim 10^{10}~\msun$. 

\section{Recovering the DM surface density profile}\label{sec:rad-prof-dm}

\begin{figure*}
\centering
\includegraphics[width=\hsize,keepaspectratio]{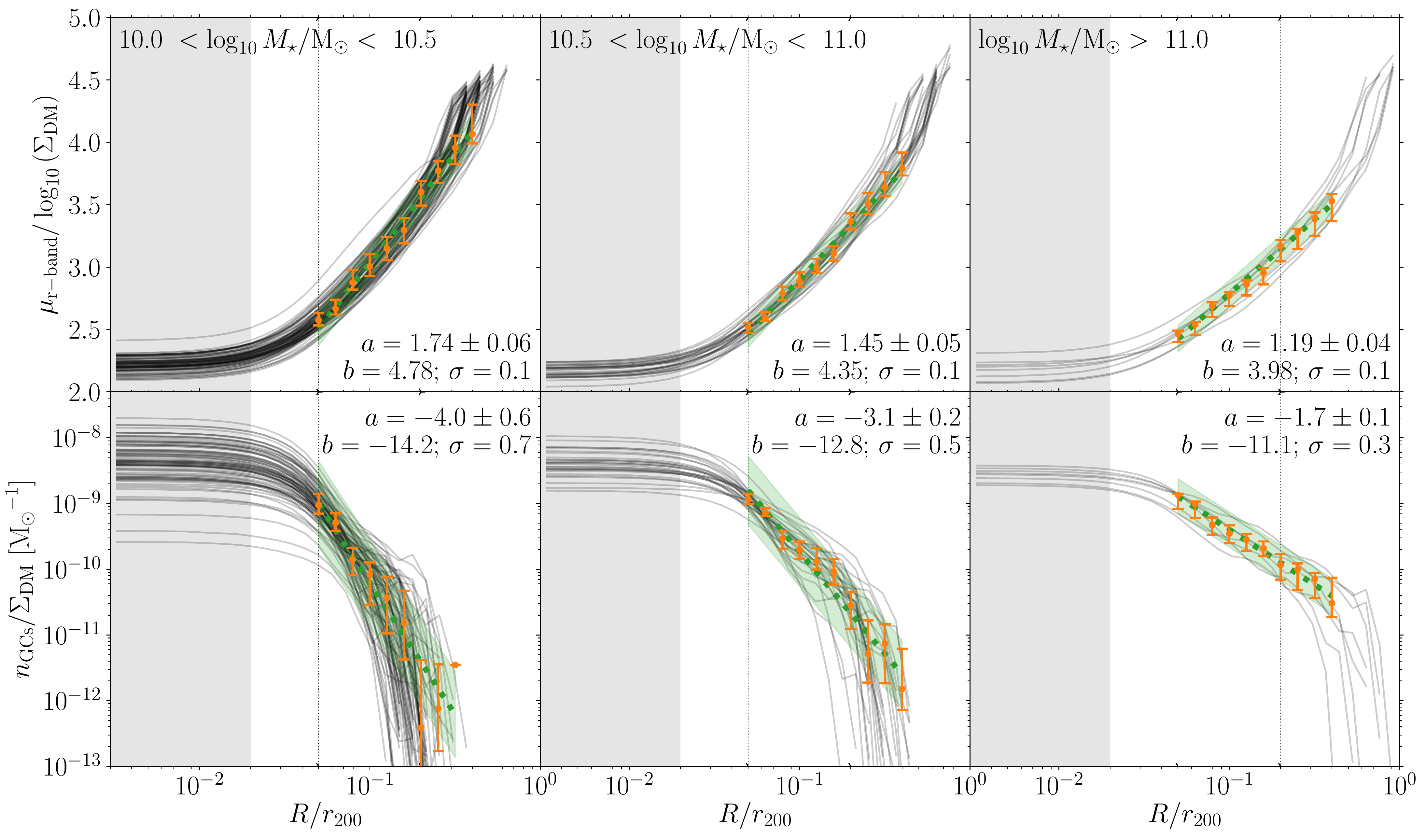}
\caption{\label{fig:ratios-stars-gcs-dm} Normalized radial profiles of the median stellar surface brightness to the DM surface density (\textit{top row}), and the median GC number density to the DM surface density (\textit{bottom row}) among our sample of haloes. The units in the upper row are $\mu_{\rm r-band}/\log_{10}\left(\Sigma_{\rm DM}/{\msun~\kpc^{-2}}\right)~[\rm mag~arcsec^{-2}]$. Each column corresponds to a different central galaxy mass bin as annotated in the top row. Thin grey lines show the individual ratios, the grey shaded region indicates the kernel size of the Gaussian filter used to smooth the images, and the thin black vertical lines mark the radial distances at which the isodensity contours are drawn. Orange markers with errorbars correspond to the median and $25$--$75$th percentiles at given radial distances. Green dotted lines with shaded regions are linear fits to the medians and the $1\,\sigma$ deviation around the fit, respectively. We indicate the coefficients of the fits and the standard deviation in each panel (see the text for details). The tight relation between these profiles implies that measuring the median stellar surface brightness or number density profile of GCs within a radial bin leads to a high-accuracy estimate of the median DM surface density in that bin ($\sigma\leq0.5~$dex, except in low-mass galaxies at large radii).} 
\end{figure*}

In this section we examine whether we can recover the DM surface density radial profile from either the stellar surface brightness maps or the number density maps of GCs.

To do this, we use the median radial profiles of stellar surface brightness, number density of GCs and DM surface density obtained during the identification of the isodensity structures (see middle panels in Fig.~\ref{fig:id-isodensity-contours-stars-gcs}). We then calculate the normalized radial profiles of the ratio of the median stellar surface brightness to the DM surface density, and of the ratio of the median number density of GCs to the DM surface density, and we show them in Fig.~\ref{fig:ratios-stars-gcs-dm}.

We find that there is a very tight relation between the median stellar surface brightness profiles and those of the DM. Similarly, there is also a tight relation between the median number density profiles of GCs and those of the DM. To quantify these relationships, we fit linear regressions of the form
\be
\mu_{\rm r-band}/\log_{10}\left(\Sigma_{\rm DM}\right) = a\log_{10}(R/r_{200}) + b
\ee
in the top row, and 
\be 
\log_{10}\left(n_{\rm GCs}/\Sigma_{\rm DM}\right) = a\log_{10}(R/r_{200}) + b
\ee
in the bottom row to the median ratios in each galaxy mass bin for the radial range $0.05 \leq R/r_{200} \leq 0.4$. We provide the best-fitting parameters in their corresponding panel, alongside the standard deviation around the linear fit. 

Across the three galaxy mass bins, the relation between the median stellar surface brightness profiles and those of the DM is tighter ($\sigma=0.1~$dex) than the relation between median number density profiles of GCs and those of the DM ($\sigma=0.3$--$0.7~$dex). The scatter around the relation between the GC number density and the DM decreases for increasing galaxy mass, suggesting that measuring the median stellar surface brightness or the median number density of GCs within a radial bin is sufficient to recover the median DM surface density within the bin with high accuracy ($\sigma\leq0.5~$dex, except in low-mass galaxies at large radii).

A similar relation between the stellar and the DM radial profiles is also found by \citet{alonsoasensio20}. Their result is based on being able to determine the average physical stellar surface density within a radial bin. In contrast, our results are based on mock observational quantities, and could thus be readily applied to observations. Hence, it opens a novel avenue to recover the DM surface density profile for a large sample of galaxies with upcoming surveys such as \textit{Euclid} or the \textit{Rubin} observatory.

\section{Projected morphology of DM, stars and GCs}\label{sec:morphology}

\begin{figure}
\centering
\includegraphics[width=\hsize,keepaspectratio]{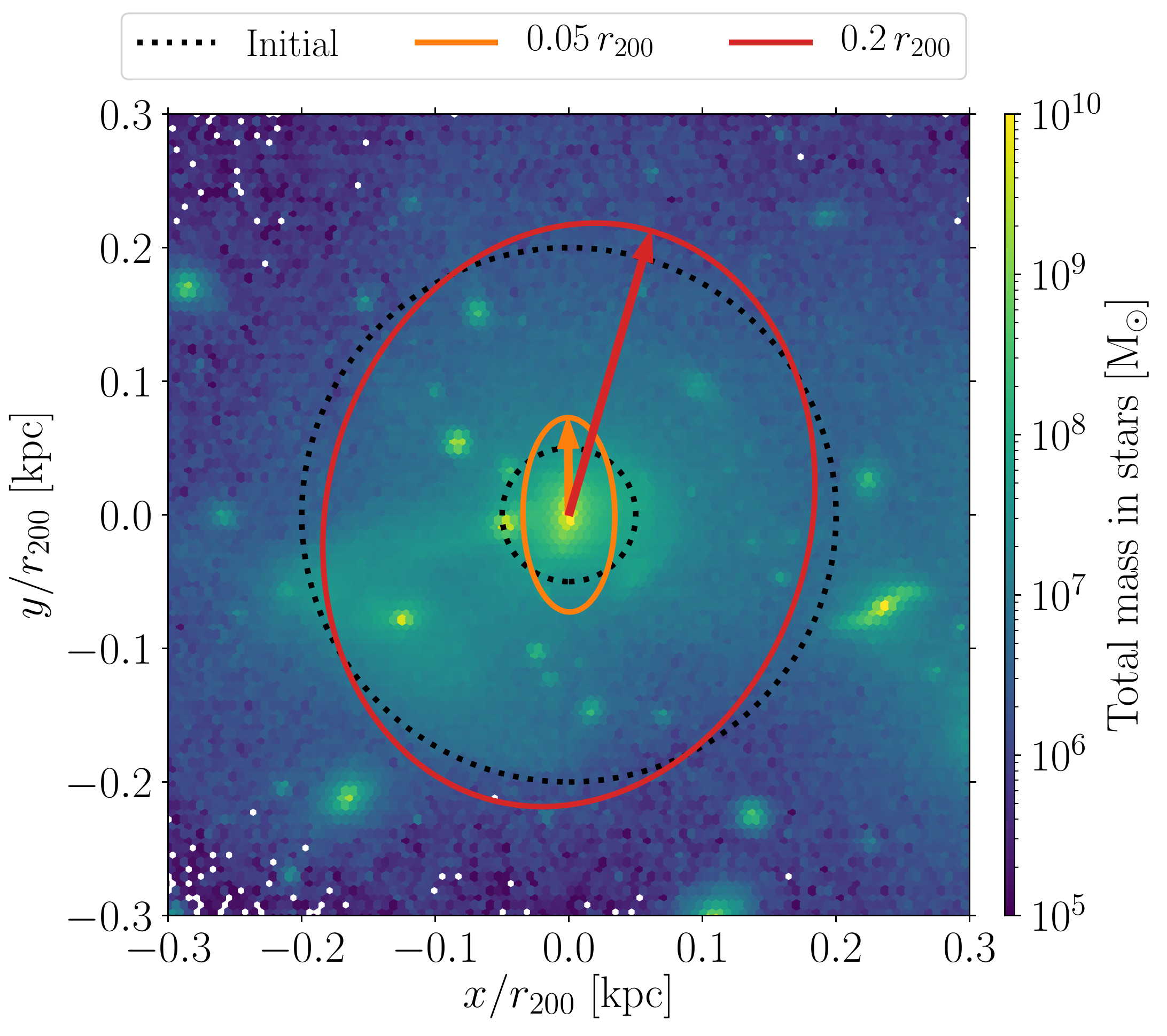}
\caption{\label{fig:shape-fitting} Example of the best-fit ellipses to the stellar distribution of FOF$000$. The black dotted lines indicate the initial apertures at the two radial distances considered. The solid lines correspond to the shape of the best-fit ellipses from the reduced mass inertia tensor, with the arrows indicating the directions of the semi-major axes.}
\end{figure}

\begin{figure}
\centering
\includegraphics[width=\hsize,keepaspectratio]{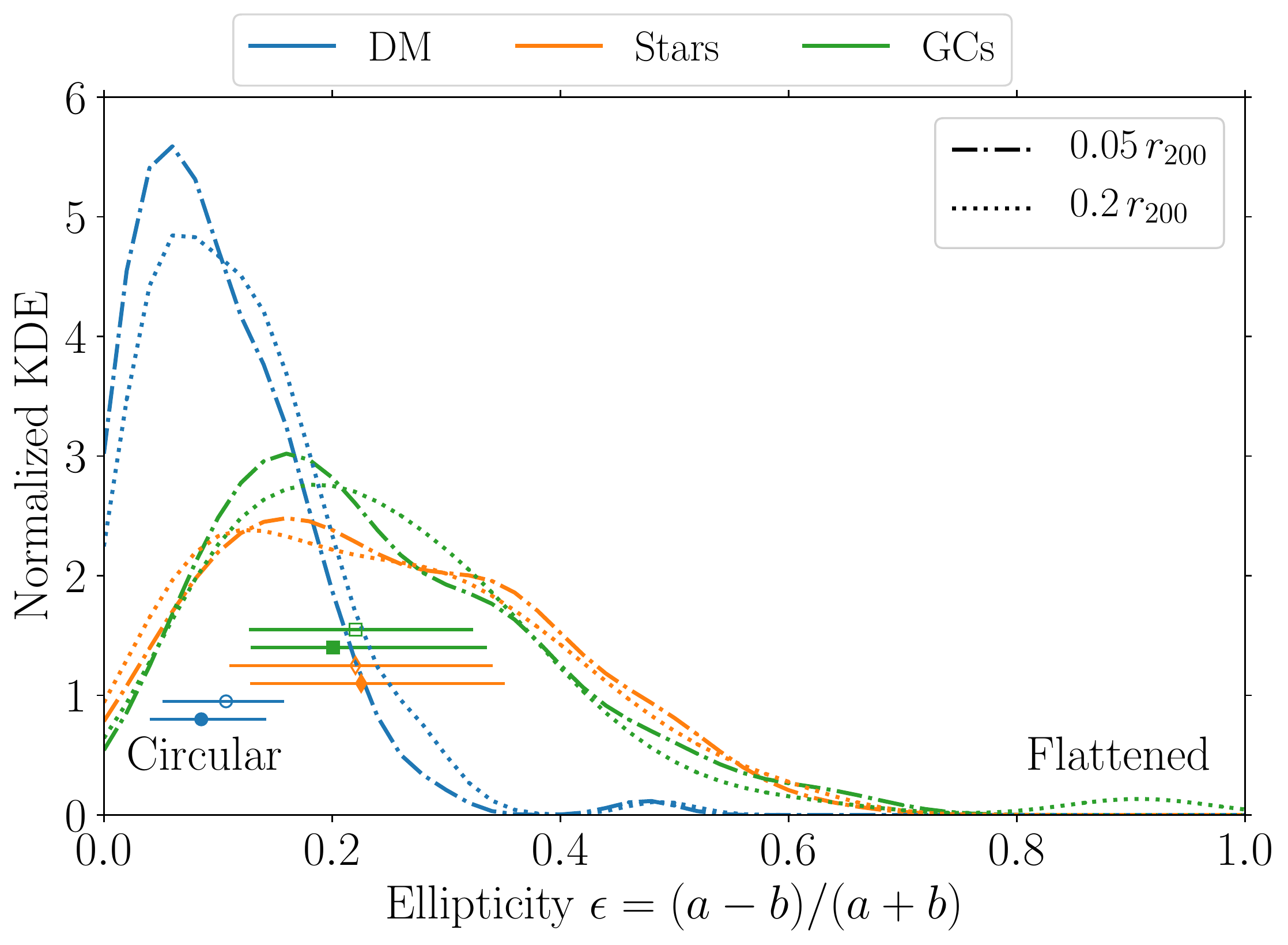}
\caption{\label{fig:hist-ellipticities} Normalized kernel density estimation of the ellipticities of the best-fit ellipses to the projected spatial distributions of DM, stars and GCs at two radial distances. Symbols with errorbars indicate the median and $25$--$75$th percentiles of each distribution. Full and empty markers correspond to the distributions for the smaller and larger initial apertures, respectively. The projected distribution of DM is circular, and the distributions of stars and GCs are slightly more flattened.}
\end{figure}

In this section, we provide quantitative descriptions of the overall  morphology of the DM, stellar and GC distributions. We do so by modelling the projected spatial distributions (as described by the particles) with ellipses. We follow the methods outlined by \cite{thob19} and \citet{hill21,hill22}\footnote{The code can be found in \href{https://github.com/Alex-Hill94/MassTensor}{https://github.com/Alex-Hill94/MassTensor}}, which we summarize below.

We focus on describing the two-dimensional spatial distributions of DM, stars and GCs with the ellipticity parameter. The ellipticity is defined as $\epsilon = (a-b)/(a+b)$, where $a$ and $b$ are the moduli of the semi-major and semi-minor axes of the ellipses, respectively. Given a particle distribution, the eigenvalues of a matrix describing its two-dimensional mass distribution ($\lambda_i$ for $i=1,2$) describe the modulus of the major and minor axes of the corresponding ellipse: $a = \sqrt{\lambda_1}$ and $b = \sqrt{\lambda_2}$ with $\lambda_1 > \lambda_2$. Similarly, the associated eigenvectors ($\mathbf{e}_{i}$ for $i=1,2$) correspond to the semi-major and semi-minor axes of the ellipse. 

We use the reduced mass distribution tensor \citep[see also][]{davis85,cole96,dubinski91,bett12,schneider12} in an iterative scheme. By considering the reduced version of the mass distribution tensor, we aim to suppress possible strong influences from features in the outer regions of the distributions, and the iterative approach enables the scheme to adapt to particle distributions that deviate from the initial selection.  The reduced inertia tensor can be described as,
\be
M_{ij}^r = \dfrac{\sum_{p} \left(m_{p}/\tilde{r}_{p}^2\right) r_{p, i} r_{p,j}}{\sum_{p} \left(m_{p}/\tilde{r}_{p}^2\right)},
\ee
where $\tilde{r}_{p}$ is the elliptical radius and $r_{p, i}$ is the $i$-th component of the coordinate vector of the particle $p$ with respect to the centre of the halo\footnote{We use the position of the particle with the lowest potential in the central galaxy to define the centre of the halo. Using instead the center of mass would cause a large shift in those haloes with prominent satellites.}. In the first iteration, the scheme considers all the particles enclosed within an initial circular aperture. For consistency with the previous sections, we quantify the morphologies within two initial apertures: $0.05\,r_{200}$ and $0.2\,r_{200}$. That provides a first estimate of the axes lengths $a$ and $b$. In the next iterations, the scheme only considers those particles that satisfy that their elliptical radius is
\be
\tilde{r}_{p}^2 \equiv \dfrac{r_{p,a}^2}{\tilde{a}^2} + \dfrac{r_{p,b}^2}{\tilde{b}^2} \leq 1,
\ee
where $\tilde{a}$ and $\tilde{b}$ are re-scaled axes lengths such that $\tilde{a}=a\, r_{p}\,(ab)^{-1/2}$. The scheme continues iterating until the fractional change in $b/a$ falls below $1$~per cent. 

Finally, we define the orientation of each component as the unit vector parallel to the major axis of the best-fit ellipse (${\mathbf{e}_1}^x$ with $x=\,$DM, stars or GCs). We determine the relative alignment between components $x$ and $y$ as the angle between these units vectors,
\be 
\theta = \arccos(|{\mathbf{e}_1}^{x} \cdot {\mathbf{e}_1}^{y}|).
\ee
The misalignment angle is invariant under a rotation of $180$~degrees, and it is thus confined between $[0,90]$~degrees. It has been shown that misalignment angles between projected distributions tend to be smaller (i.e.~more aligned) at all radii and halo masses than when measured in three dimensions \citep{tenneti14,velliscig15a}. We show an example of the best-fit ellipses to the stellar distribution of FOF$000$ in Fig.~\ref{fig:shape-fitting}.

We thus quantify the morphology of the projected spatial distributions of DM, stars and GCs in our sample of $117$ haloes using all the particles in their host halo as determined by the \fof algorithm. A strong constraint of the iterative scheme in order to obtain convergence is to have at least $100$ particles within the initial circular aperture. Given the small sizes of the GC populations, this requirement decreases the number of haloes for which we can apply the scheme to characterise the morphology of GCs. We can do so to 88 and 93 haloes, respectively, for the $0.05\,r_{200}$ and $0.2\,r_{200}$ initial apertures.

\begin{table}
\centering{
  \caption{Characterising the morphology and relative alignment between matter components. From left to right, columns indicate the initial aperture, the ellipticities of the DM, stellar and GC distributions at the top, as well as the misalignment angles between the different components at the bottom. Values correspond to the medians, and we indicate the $25$--$75$th percentiles in parentheses.}
  \label{tab:median-misalignment-angles-ellipticities}
	\begin{tabular}{cccc}
	\hline\hline
		\multicolumn{4}{c}{Ellipticities}\\ 
		Aperture & DM & Stars & GCs \\ \hline
		$0.05\,r_{200}$ & $0.09$ ($0.04$--$0.14$) & $0.23$ ($0.13$--$0.35$) & $0.20$ ($0.13$--$0.34$) \\
		$0.2\,r_{200}$ & $0.11$ ($0.05$--$0.16$) & $0.22$ ($0.11$--$0.34$) & $0.22$ ($0.13$--$0.32$) \\ 
        \hline \hline
		\multicolumn{4}{c}{Misalignment angles [deg]}\\ 
		Aperture & GCs -- DM & GCs -- Stars & Stars -- DM \\ \hline
		$0.05\,r_{200}$ & $9.48$ ($4.27$--$21.27$) & $7.94$ ($3.74$--$19.27$) & $6.96$ ($2.01$--$19.49$)\\
		$0.2\,r_{200}$ & $11.23$ ($3.10$--$29.17$) & $9.79$ ($3.59$--$15.70$) & $7.99$ ($2.03$--$23.32$) \\ 
		\hline \hline
	\end{tabular}}
\end{table}

We show the distribution of the ellipticities of the best-fit ellipses to the projected spatial distributions of DM, stars and GCs in Fig.~\ref{fig:hist-ellipticities}. Overall, we find that the DM distributions are close to being circular, whereas the stellar and GC distributions are more flattened. For the smaller initial aperture ($0.05\,r_{200}$), the median ellipticity of DM is $0.09$, whereas the median ellipticities of stars and GCs are $0.23$ and $0.20$, respectively. The distribution of ellipticities of the stellar and GC distributions are very similar, with the $25$--$75$th percentiles spanning the same range of values, $\sim0.13$--$0.35$. When considering the outer initial aperture ($0.2\,r_{200}$), we find that the distribution of ellipticities of the stellar and GC populations barely change. We relate this to most of their mass already being enclosed within the smaller initial aperture. In contrast, the DM distributions appear slightly more flattened with a median of $0.11$. The flattened stellar and GC distributions might be explained by the flatness of galaxy disks, which reside within $r<0.05\,r_{200}$, as a result of the dissipative collapse of the gas. One way to improve the agreement between the DM and the stellar and GC distributions would be to remove the central galaxy from the distribution being fitted. Another method would be to remove the mass weighting in the reduced inertia tensor, as most of the mass in stars and GCs mass resides in the inner region of the halo. We summarize the median and the $25$--$75$th percentiles of the ellipticity distributions of DM, stars and GCs in Table~\ref{tab:median-misalignment-angles-ellipticities}.

\begin{figure}
\centering
\includegraphics[width=\hsize,keepaspectratio]{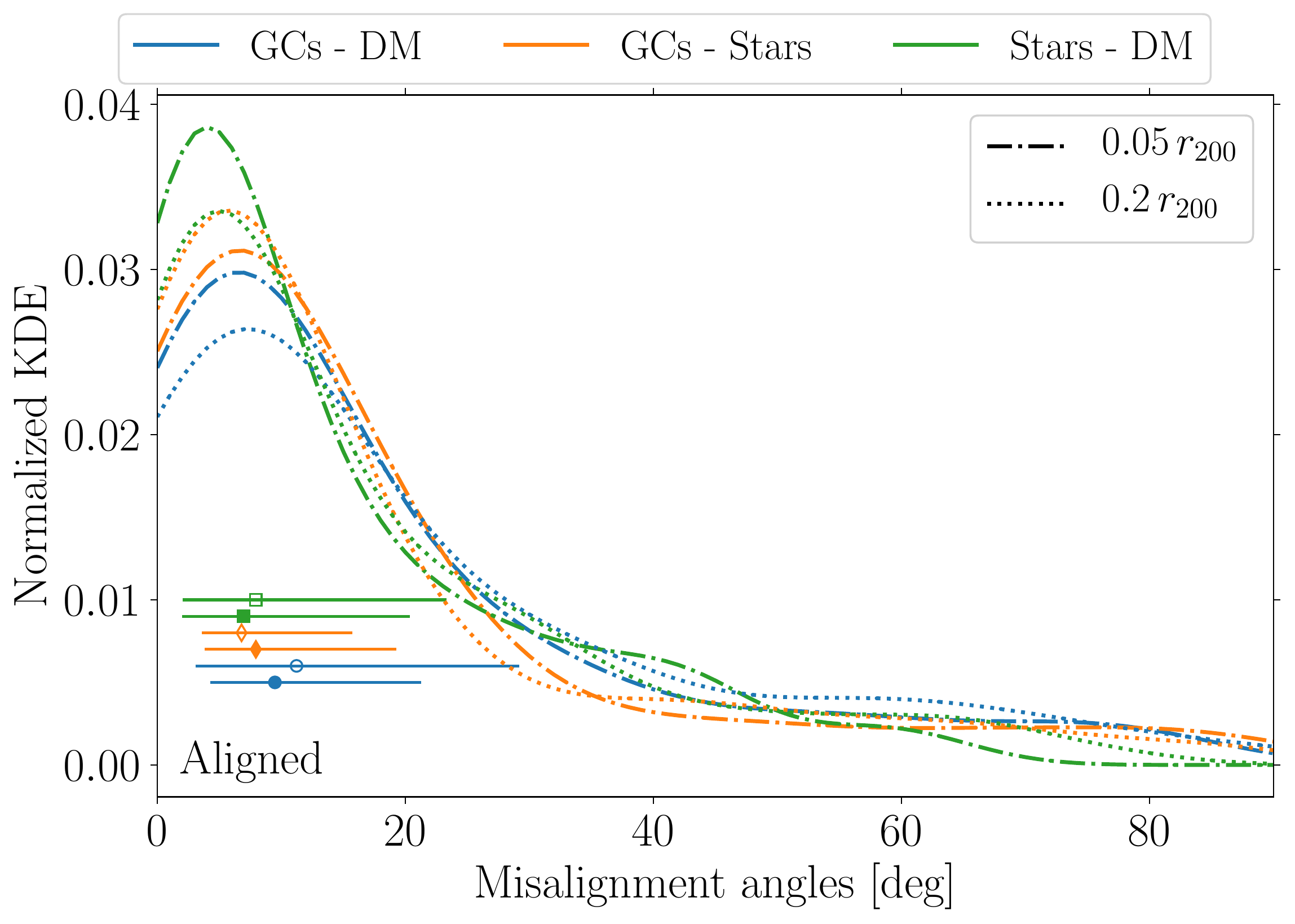}
\caption{\label{fig:hist-misalignment-angles} Normalized kernel density estimation of the relative alignment angles between the semi-major axes of the projected spatial distributions of DM, stars and GCs at two radial distances. Symbols with errorbars indicate the median and $25$--$75$th percentiles of each distribution. Full and empty markers correspond to the distributions for the smaller and larger initial apertures, respectively. All spatial distributions are preferentially aligned, with stars and DM showing the closest alignment.}
\end{figure}

We also examine the relative alignment between the semi-major axes of the best-fit ellipses to the DM, stellar and GC distributions, which we show in Fig.~\ref{fig:hist-misalignment-angles}. Regardless of the initial aperture, we find that all the distributions are preferentially aligned, with the orientation of $\sim 70~$per cent of the GC systems being less than $20~$degrees away from their host DM halo. The DM and stellar distributions show the closest alignment with a median difference of $\sim8$--$9~$degrees at the two radial distances considered \citep[also see fig.~13 in][]{hill21}. GCs are more aligned with the stellar than with the DM distribution of their host halo, although the median misalignment angles among all components only differ by $\sim 2$--$3$ degrees. Previous studies have already found that galaxies in the \eagle simulations are well aligned with the local mass distribution, but are often misaligned with respect to the global halo. This indicates that the stellar component traces the local DM distribution, which is the dominant matter component, and that the structure of the DM halo changes from the inner to the outer halo \citep{velliscig15a,hill21}.  

It is worth noting that some severe misalignments are simply cases where the two components are actually aligned, just not along the same axes. This happens most often in very prolate haloes \citep[e.g.~see fig.~7 of][]{hill21}. We have also examined the ellipticities and the misalignment for different galaxy stellar mass bins, but no trends could be identified. Thus, we find that characterising the orientation of the spatial distribution of stars and GCs is a very accurate probe of the orientation of the DM halo.

\section{Conclusions}\label{sec:conclusions}

In this work, we explore whether or not diffuse stellar and GC populations trace the overall structure of the dark and total matter distribution of their host halo. For this, we use the $(34.4~\rm cMpc)^3$ periodic volume from the \emosaics project \citep{pfeffer18,kruijssen19a}. The combination of a sub-grid description for stellar cluster formation and evolution with the \eagle galaxy formation model \citep{schaye15,crain15} allows us to model the simultaneous formation and assembly of stellar cluster populations and their host galaxies over a Hubble time. 

We explore the diffuse stellar light and GCs populations hosted by the $117$ DM haloes in our simulation containing central galaxies more massive than $M_{\star}>10^{10}~\msun$, which corresponds to halo masses $M_{200}\gtrsim 4\times 10^{11}~\msun$. For each of these haloes, we create projected surface density maps of DM, total mass (i.e.~the addition of the DM, gas and stellar surface density), as well as stellar surface brightness maps and projected number density maps for the GC populations (Sect.~\ref{sub:maps}). To remove small-scale noise, we smooth the maps with a Gaussian kernel of size $\sigma=0.02\,r_{200}$ (Fig.~\ref{fig:proj-dm-totmass-stars-gcs-smoothed}), which roughly corresponds to the effective radius of the central galaxy \citep{kravtsov13}. We demonstrate in App.~\ref{app:gaussian-filter} the influence of changing the size of the kernel on the structures that can be identified in each map. 

To determine the similarity between the maps, we begin by calculating the  Pearson correlation coefficient between the same pixels in pairs of smoothed maps (Fig.~\ref{fig:pearsoncoeff-mstar}). We find that both the stellar surface brightness maps and the projected number density maps of GCs are uncorrelated with randomized fields, indicating that there are coherent structures in the images. The stellar light shows the highest degree of correlation to the DM distribution across our sample, whereas GC populations hosted by lower mass systems are increasingly uncorrelated to both the stellar and the DM distributions.

We then identify structures in our maps by drawing isodensity contours at the corresponding average densities for a given set of radial distances (Fig.~\ref{fig:id-isodensity-contours-stars-gcs}). For each radial distance, we keep the longest contour as the dominant structure. In haloes with prominent satellites (e.g.~ongoing major mergers), the algorithm can identify different structures in different tracers (i.e.~the central and the satellite), and thus the comparison of the contours becomes more challenging. By comparing the identified structures in each map, we find that, in most of our haloes, the structures identified in the diffuse stellar light and in the GC populations agree with those in the DM and total mass maps (Fig.~\ref{fig:overplot-dmmaps-contours}).

We quantify the comparison by calculating the Modified Hausdorff Distance (MHD), which is a measure of the difference between pairs of contours (Fig.~\ref{fig:mhd-r200-mstar}). The MHD provides a quantitative measure of the accuracy of the reconstruction of the DM contours using GCs and stars. We consider the surface density maps of DM and total mass as the \emph{true} distributions that we aim to recover, and the stellar surface brightness maps and the number density maps of GCs as the \emph{observational} distributions. For our fiducial limits ($\mu_{r-{\rm band}} < 28~\rm mag/arcsec^{2}$ in the stellar maps and $M_{F475W}<-6.217$ for the GCs), we find that the GC populations are better tracers of the matter distribution in the inner region of the halo ($0.05\,r_{200}$), whereas the diffuse stellar light is more accurate at larger radii. However, observing GC populations is less challenging than mapping the diffuse stellar light, and thus it can be a more effective way of probing their host DM halo.

By examining the radial profile of the accreted fractions of stars and GCs (Fig.~\ref{fig:accreted-fraction}), we find that the accuracy of each tracer is related to the dominance of accreted objects, i.e.~the material deposited during the assembly of the host halo contains the information about the matter distribution. At the two radial distances considered, the fraction of accreted GCs is greater than or similar to that of stars. In low mass haloes ($10 <\log_{10}(M_{\star}/\msun) < 10.5 $), the accreted and in-situ fractions become comparable, and thus the signal tracing the matter distribution dilutes and it is more difficult to recover.

As a proof-of-concept, we examine how much the measured MHDs relative to the DM isodensity contours change as a result of three different effects. Firstly, we examine the influence of the presence of substructure within each halo. We find that removing the galactic substructure from our maps has an effect smaller than $0.01\,r_{200}$ in the measured MHDs relative to the DM contours (Fig.~\ref{fig:compare-mhd-r200-mstar}), and that there are only four instances for which it introduces a significant deviation. This is a promising result for future applications of this analysis to  observational data since it is more difficult to remove substructure in observations.

Secondly, we consider the effect of varying the stellar surface brightness limit applied to the maps (Fig.~\ref{fig:compare-mhd-r200-mstar-surfbrightlim}). We consider the limits from \textit{SDSS}, the \textit{Euclid} Wide Survey and the $10$-year integration limit of the \textit{Rubin} observatory, and compare the results to our fiducial limit. We find that deeper observations do not lead to smaller MHDs, but allow us to apply this analysis to lower-mass haloes and to larger radial distances within the halo.

Thirdly, we examine the role of the GC absolute magnitude limit by selecting GC populations that reach above and below the fiducial value (i.e.~the peak of the GC luminosity function, Fig.~\ref{fig:compare-mhd-r200-mstar-maglim}). We find that observing fainter GCs also allows this analysis to be extended to lower-mass haloes and to larger radial distances within the halo. Additionally, we find that the contours identified in the deeper GC observations are closer to the DM isodensity contours than those identified in brighter populations. 

As a next step, we examine whether the DM surface density radial profile can be recovered from either the stellar surface brightness maps or the number density maps of GCs (Fig.~\ref{fig:ratios-stars-gcs-dm}). We find that there is a very tight relation between the median stellar surface brightness profiles and those of DM (the scatter around the fit is $\sigma=0.1~$dex). Albeit with more scatter ($\sigma=0.3$--$0.7~$dex), there is also a tight relation between the median number density profiles of GCs and those of DM. We fit linear regressions to the median rations for the radial range $0.05 \leq R/r_{200} \leq 0.4$, and provide the coefficients in Fig.~\ref{fig:ratios-stars-gcs-dm}. Thus, measuring the median stellar surface brightness or the median number density of GCs within a radial bin is sufficient to recover the median DM surface density within the bin with high accuracy ($\sigma\leq0.5~$dex, except for low-mass galaxies at large radii). Compared to the relation provided by \citet{alonsoasensio20}, our result is based on mock observations. Thus, it opens a novel avenue to recover the DM surface density profile for a large sample of galaxies with upcoming surveys with the \textit{Euclid} or the \textit{Rubin} observatories.

We then quantify the projected morphology of DM, stars and GCs by modelling the spatial distribution of their particles using ellipses \citep[also see e.g.][]{tenneti14,velliscig15a,thob19,hill21}. We perform the fitting technique for each halo in our sample considering two initial circular apertures at $0.05\,r_{200}$ and at $0.2\,r_{200}$. We find that DM shows the most circular distributions, whereas those of stars and GCs are more flattened (Fig.~\ref{fig:hist-ellipticities}). Considering a larger initial aperture has a negligible effect on the ellipticities measured.

We also examine the relative alignments between the semi-major axes of the best-fitting ellipses describing the DM, stellar and GC populations in our sample (Fig.~\ref{fig:hist-misalignment-angles}). We find that all distributions are preferentially aligned, with median misalignment values of $\sim 10~$degrees and the $75$th percentiles being less than $30~$degrees (Table~\ref{tab:median-misalignment-angles-ellipticities}). The stellar distribution shows a closer alignment to DM than GCs, but the difference is only $\sim 2~$degrees. Stars have previously been shown to be a good tracer of the local matter distribution \citep{velliscig15a,hill21}, and our results suggest GC populations are also effective tracers. Thus, by characterising the orientation of diffuse stellar light and GC populations we can determine the orientation of their host DM halo within a few degrees.

Our results suggest that the host DM haloes of observed galaxies can be mapped out to distances of nearly half the virial radius using the projected distribution of GCs, and to even further distances using very deep stellar observations. The accuracy when using GCs is highest at $0.05\,r_{200}$, and for galaxies with the most abundant GC systems (including massive ellipticals). In the local Universe ($D < 50~\mpc$), observations of GC populations do not require deep observations, in contrast with diffuse stellar haloes. Therefore, maps of GC systems provide a more efficient way to trace the DM distribution. Upcoming deep and wide surveys with the \textit{Euclid}, \textit{Roman}, and \textit{Rubin} observatories will allow the application of this method to reconstruct the host DM haloes of very large samples of galaxies. Comparing those results to DM maps of hundreds of galaxy clusters obtained using strong lensing analyses in next-generation surveys could yield new constraints on the nature of DM.

\section*{Acknowledgements}
MRC thanks Bill Harris for valuable discussions on mimicking observations, and Alex Hill for providing his code to measure the shape of particle distributions.
MRC gratefully acknowledges the Canadian Institute for Theoretical Astrophysics (CITA) National Fellowship for partial support, and this work was supported by the Natural Sciences and Engineering Research Council of Canada (NSERC). STG and JMDK gratefully acknowledge funding from the European Research Council (ERC) under the European Union's Horizon 2020 research and innovation programme via the ERC Starting Grant MUSTANG (grant agreement number 714907).
JLP is supported by the Australian government through the Australian Research Council’s Discovery Projects funding scheme (DP200102574).
AD is supported by a Royal Society University Research Fellowship. AD also acknowledges support from the Leverhulme Trust and the Science and Technology Facilities Council (STFC) [grant numbers ST/P000541/1, ST/T000244/1]. 
RAC is a Royal Society University Research Fellow.
JMDK gratefully acknowledges funding from the German Research Foundation (DFG) in the form of an Emmy Noether Research Group (grant number KR4801/1-1). 

This work used the DiRAC Data Centric system at Durham University, operated by the Institute for Computational Cosmology on behalf of the STFC DiRAC HPC Facility (www.dirac.ac.uk). This equipment was funded by BIS National E-infrastructure capital grant ST/K00042X/1, STFC capital grants ST/H008519/1 and ST/K00087X/1, STFC DiRAC Operations grant ST/K003267/1 and Durham University. DiRAC is part of the National E-Infrastructure. The work also made use of high performance computing facilities at Liverpool John Moores University, partly funded by the Royal Society and LJMU's Faculty of Engineering and Technology.

\textit{Software}: This work made use of the following \code{Python} packages: \code{h5py} \citep{h5py_allversions}, \code{Jupyter Notebooks} \citep{kluyver16}, \code{Numpy} \citep{numpy-harris20}, \code{Pynbody} \citep{pynbody} and \code{Scipy} \citep{jones01}, and all figures have been produced with the library \code{Matplotlib} \citep{hunter07}. The comparison to observational data was done more reliably with the help of the WEBPLOT- \code{DIGITIZER}\footnote{\href{https://apps.automeris.io/wpd/}{https://apps.automeris.io/wpd/}} webtool.

\section*{Data Availability}

The data underlying this article will be shared on reasonable request to the corresponding author.



\bibliographystyle{mnras}
\interlinepenalty=10000 
\bibliography{bibdesk-bib}



\appendix

\section{Influence of the kernel size}\label{app:gaussian-filter}

\begin{figure*}
\centering
\includegraphics[width=\hsize,keepaspectratio]{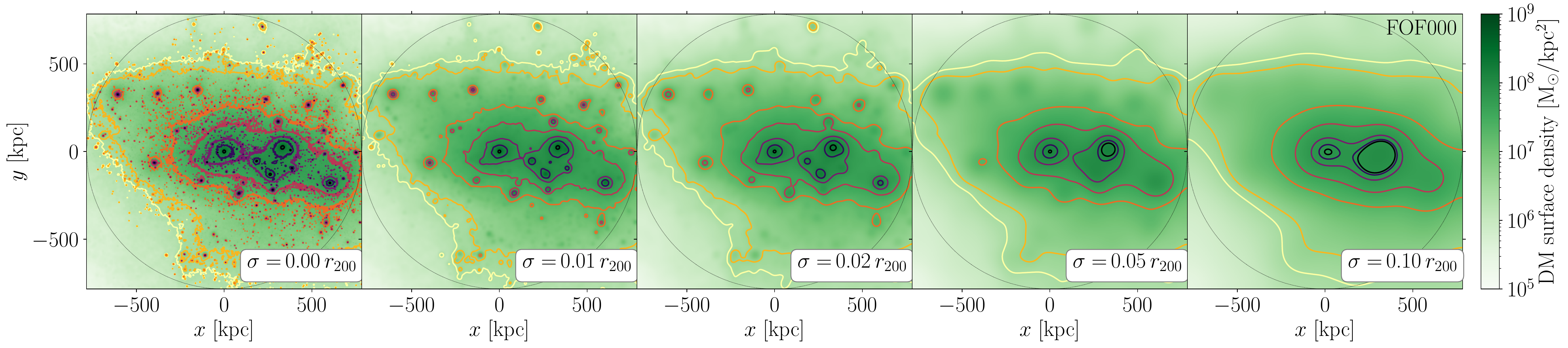}
\includegraphics[width=\hsize,keepaspectratio]{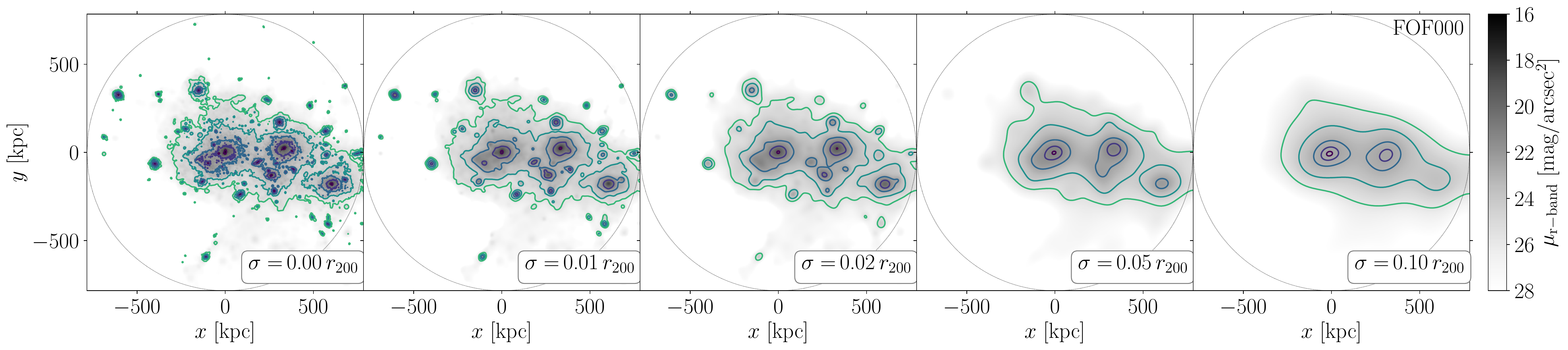}
\includegraphics[width=\hsize,keepaspectratio]{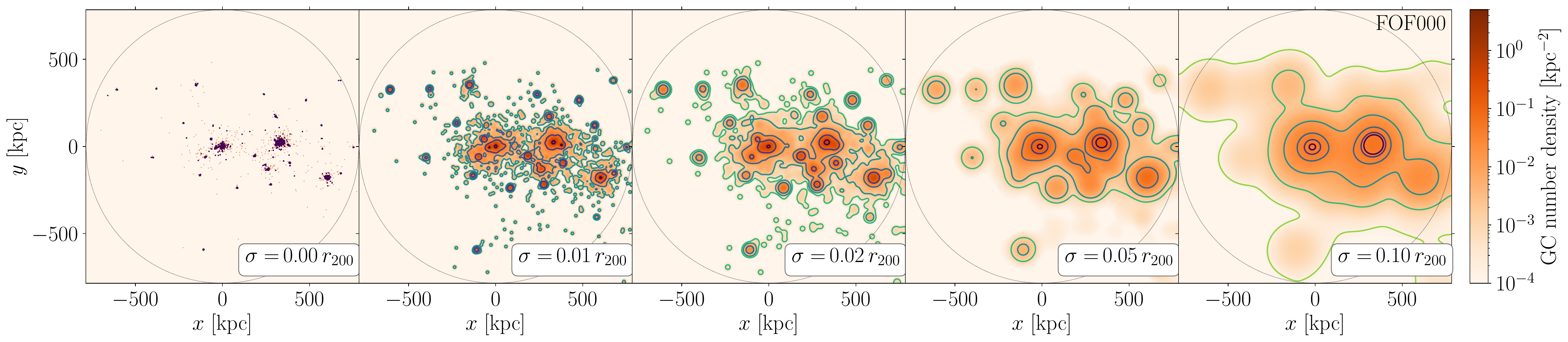}
\caption{\label{fig:app-kernel-size-contours-stars-gcs}  Comparison of the projected stellar surface density maps of DM (\textit{top row}), the stellar surface brightness maps (\textit{middle row}) and the number density maps of GCs (\textit{bottom row}) hosted by FOF$000$ and smoothed with Gaussian kernels of different sizes. From left to right, images have been smoothed by kernels of sizes $\sigma=0,0.01,0.02,0.05,0.10\,r_{200}$, with the former indicating that no smoothing was applied. The isodensity contours are identified as described in Sect.~\ref{sub:isodensity-contours} and Fig.~\ref{fig:id-isodensity-contours-stars-gcs}. As a reference, the thin dotted black circle marks the extent of the virial radius of the halo.}
\end{figure*}

In this Appendix we explore the influence of the size of the Gaussian kernel on the structures that can be identified from the smoothed maps. We consider five different kernel sizes, $\sigma=0,0.01,0.02,0.05,0.10\,r_{200}$, with the former implying that no smoothing is performed. We show the resulting smoothed DM surface density, stellar surface brightness and GC number density maps of FOF$000$ in Fig.~\ref{fig:app-kernel-size-contours-stars-gcs}. As we increase the size of the kernel, we find that the small scale noise is smoothed out. Once the kernel is larger than the effective radius of the central galaxy \citep[$\sim0.015\,r_{200}$;][]{kravtsov13}, the galactic-scale signal is washed out and the isodensity contours are very smooth. For these reasons, we decide to use a kernel size of $\sigma=0.02\,r_{200}$ in our main analysis.

\begin{figure*}
\centering
\includegraphics[width=\hsize,keepaspectratio]{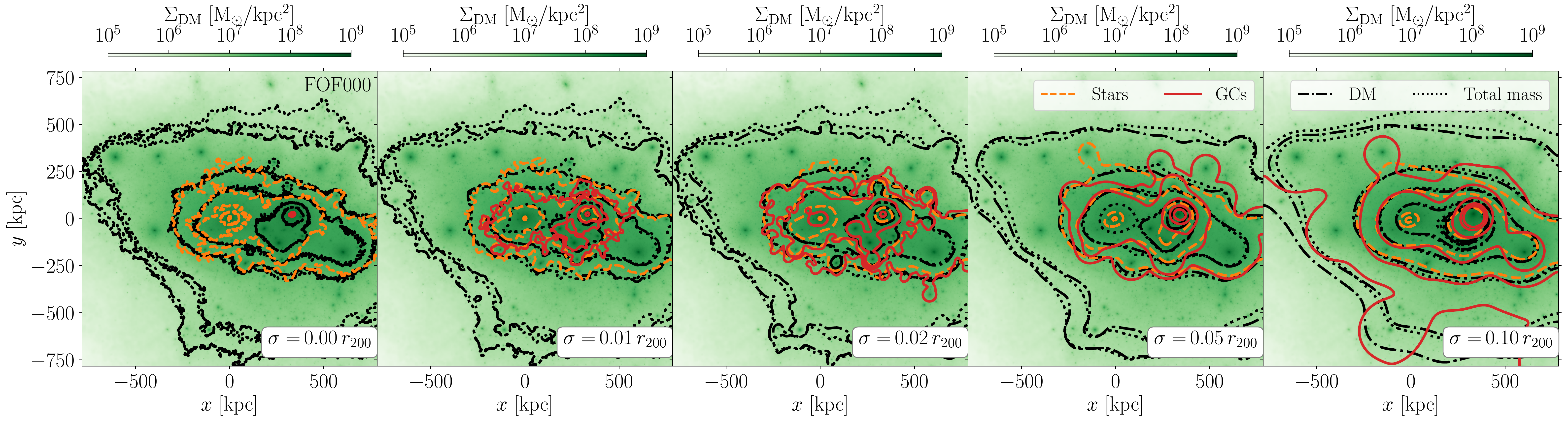}
\caption{\label{fig:app-kernel-size-overplot-contours} Qualitative comparison of the structures identified in the surface density maps of DM and total mass (black lines), and on the stellar surface brightness maps (dashed orange lines) and the number density maps of GCs (solid red lines) hosted by halo FOF$000$. From left to right, the isodensity contours are identified on projected images smoothed with Gaussian kernels of sizes $\sigma=0,0.01,0.02,0.05,0.10\,r_{200}$, with the former indicating that no smoothing was applied. The background image corresponds to the unsmoothed surface density map of DM in FOF$000$.}
\end{figure*}

We compare the dominant isodensity contours identified in the maps of the different components smoothed with Gaussian kernels of different sizes in Fig.~\ref{fig:app-kernel-size-overplot-contours}. Using a kernel of size $\sigma=0.02\,r_{200}$, we smooth out the small scale noise while retaining the galactic-scale perturbations in the contours. 

\bsp	
\label{lastpage}
\end{document}